\documentstyle[astrobib,psfig]{mn-ab}
\newcommand{\lcdm} {\mbox{$\Lambda$CDM}}
\newcommand{\taucdm} {\mbox{$\tau$CDM}}
\newcommand{\hkpc}{\mbox{$h^{-1}$ kpc}}
\newcommand{\hmpc}{\mbox{$h^{-1}$ Mpc}}
\newcommand{\hmsun}{\mbox{$h^{-1}$ $M_{\odot}$}}

\newcommand{\kms}{\mbox{km s$^{-1}$}}

\newcommand{\msun}{\mbox{$M_{\odot}$}}

\newcommand{\kmsmpc}{\mbox{km s$^{-1}$ Mpc$^{-1}$}}

\def\la{\mathrel{\hbox{\rlap{\hbox{\lower4pt\hbox{$\sim$}}}\hbox{$<$}}}}
\def\ga{\mathrel{\hbox{\rlap{\hbox{\lower4pt\hbox{$\sim$}}}\hbox{$>$}}}}

\def\be{\begin{equation}}  \def\ee{\end{equation}}
\def\av#1{\langle #1 \rangle}
 
\def\delg{\delta_{\rm g}}

\def\sigmab{\sigma_{\rm b}}
\def\bxi{b_\xi}
\def\bvar{b_{\rm var}}
\def\bhat{\hat{b}}
\def\b1{b_{1}}
\def\Rs{R_{\rm s}}
\def\ltsima{$\; \buildrel < \over \sim \;$}
\def\lsim{\lower.5ex\hbox{\ltsima}}
\def\gtsima{$\; \buildrel > \over \sim \;$}
\def\gsim{\lower.5ex\hbox{\gtsima}}
\def\ga{\mathrel{\hbox{\rlap{\hbox{\lower4pt\hbox{$\sim$}}}\hbox{$>$}}}}
\def\la{\mathrel{\hbox{\rlap{\hbox{\lower4pt\hbox{$\sim$}}}\hbox{$<$}}}}

\title{Non-linear Stochastic Galaxy Biasing in Cosmological Simulations}      
\author[R.S. Somerville, G. Lemson, Y. Sigad, A. Dekel,
       G. Kauffmann \& S.D.M. White]
       {Rachel S. Somerville$^1$, Gerard Lemson$^{1,2}$,
        Yair Sigad$^1$, Avishai Dekel$^{1}$, 
	\newauthor
	Guinevere Kauffmann$^2$ \& Simon D.M. White$^2$ \\
        $^1$Racah Institute of Physics, The Hebrew University, Jerusalem\\
        $^2$Max-Planck Institut f\"{u}r Astrophysik, D-85740 Garching, 
	Germany}

\begin{document}

\maketitle

\begin{abstract}
We study the biasing relation between dark-matter halos or galaxies
and the underlying mass distribution, using cosmological $N$-body
simulations in which galaxies are modelled via semi-analytic recipes.
The nonlinear, stochastic biasing is quantified in terms of the mean
biasing function and the scatter about it as a function of time, scale
and object properties. The biasing of galaxies and halos shows a
general similarity and a characteristic shape, with no galaxies in
deep voids and a steep slope in moderately underdense regions. At
$\sim 8\hmpc$, the nonlinearity is typically $\lsim 10$ percent and
the stochasticity is a few tens of percent, corresponding to $\sim 30$
percent variations in the cosmological parameter
$\beta=\Omega^{0.6}/b$. Biasing depends weakly on halo mass, galaxy
luminosity, and scale. The observed trend with luminosity is
reproduced when dust extinction is included. The time evolution is
rapid, with the mean biasing larger by a factor of a few at $z\sim 3$
compared to $z=0$, and with a minimum for the nonlinearity and
stochasticity at an intermediate redshift. Biasing today is a weak
function of the cosmological model, reflecting the weak dependence on
the power-spectrum shape, but the time evolution is more
cosmology-dependent, relecting the effect of the growth rate. We
provide predictions for the relative biasing of galaxies of different
type and color, to be compared with upcoming large redshift surveys.
Analytic models in which the number of objects is conserved
underestimate the evolution of biasing, while models that explicitly
account for merging provide a good description of the biasing of halos
and its evolution, suggesting that merging is a crucial element in the
evolution of biasing.
\end{abstract}

\section{Introduction}
\label{sec:intro}
The standard picture of the growth of structure via gravitational
instability within a dark-matter dominated universe has led to a
successful and predictive theoretical framework. However, in order to
make direct contact with most observations, the relationship between
galaxies and the underlying dark-matter distribution must be
understood. This relationship has come to be loosely referred to as
galaxy ``biasing'' \cite{davis:85,bbks,dekel-rees:87}.

Given the complexity of the process of galaxy formation, it would be
surprising if the galaxy distribution traced the mass distribution in
a simple way. Various physical mechanisms have been proposed that
could lead to galaxy biasing, but the features of the biasing process
remain highly uncertain, including, for example, non-linearity, scale
dependence, and stochasticity.  Even the direction of the biasing,
i.e. bias or anti-bias, is sometimes uncertain.

Yet, there are many observational indications that galaxy biasing does
exist and is non-trivial; for example, the dependence of galaxy
clustering on type or environment
\cite{dressler:80,hermit:96,willmer:98,tegmark-bromley:98}, or the
evolution of biasing in time \cite{steidel:spike,adelberger:98}. To
the extent that the cosmological parameters are known, studying galaxy
biasing will allow us to better understand the process of galaxy
formation. Conversely, better understanding of biasing is a crucial
component in interpreting the results of methods that attempt to use
galaxies as tracers of the underlying density field and thus determine
the cosmological density parameter $\Omega$ (see reviews by
\citeNP{strauss-willick:95}, \citeNP{dekel:94}, \citeNP{dekel:97}, and
\citeNP{dekel:99}).

Modern $N$-body techniques have provided the basis for a great deal of
progress in understanding the clustering of dark-matter halos relative
to the underlying matter density field (e.g. \citeNP{gross:98},
\citeNP{jenkins:98}). Analytic approximations based within the
hierarchical-merger framework have led to additional insights
(e.g. \citeNP{mw} and extensions, cf. \citeNP{catelan:98}). However,
these halos do not necessarily correspond to galaxies. One reason for
this is the ``over-merging'' problem: dark matter halos that are
incorporated within a larger collapsed structure may lose their
identities due to the limited numerical resolution of the
simulation. In addition, the biasing properties of the halos include
only the effects of clustering due to gravity and do not reflect the
astrophysical processes involving gas and stars, which are presumably
important in determining the properties of real galaxies.

Recent theoretical studies have used a variety of techniques to
address these issues. For example, \citeN{cen-ostriker:98},
\citeN{katz:98} \citeN{blanton:98} and \citeN{blanton:99} used
cosmological simulations with hydrodynamics and phenomenological
recipes for star formation. \citeN{colin:98} used very high-resolution
dissipationless N-body simulations, in which the over-merging problem
is largely overcome and there should be a good one-to-one
correspondence between dark matter halos and
galaxies. \citeN{narayanan:98} and \citeN{mann:98} used various ``toy
models'' to describe galaxy formation in large, low resolution
dissipationless N-body simulations. \citeN{kns} and \citeN{benson:99}
used semi-analytic techniques to assign galaxies to halos within
dissipationless N-body simulations.

All of these techniques have certain advantages and disadvantages.
The present study provides a useful complement to this previous work,
and is unique in several respects. We model galaxies using a new
technique \cite{kauffmann:98a}, in which $N$-body simulations are
combined with semi-analytic modelling of galaxy formation. Merging
histories of dark matter halos (hereafter referred to simply as halos)
are extracted from large dissipationless N-body simulations using the
outputs at finely spaced time steps. Within the framework of these
``merger trees'', gas dynamics, star formation, supernovae feedback,
and galaxy merging are modelled semi-analytically. The results are
convolved with stellar population models. In this way, we obtain
predictions of observable galaxy properties such as magnitudes,
colours, star formation rates, and morphologies. This allows us to
study the dependence of galaxy biasing on these characteristics at
different cosmological epochs and at different smoothing scales.
Another novel aspect of this work is that we utilize the biasing
formalism developed by \citeN{dekel-lahav}, which allows us to
separate and quantify the nonlinearity and the stochasticity of the
biasing relation.

The outline of the paper is as follows. In
Section~\ref{sec:formalism}, we summarize the biasing formalism and
analytic models for halo biasing. In Section~\ref{sec:sims}, we
briefly describe the simulations and the semi-analytic techniques used
to model galaxy formation. In Section~\ref{sec:masstolight}, we
discuss the relationship between the halos and galaxies in our
simulations. In Section~\ref{sec:results}, we study the dependence of
the biasing relation on halo mass, galaxy luminosity and type, scale,
and redshift. In Section~\ref{sec:analytic}, we compare the results of
the simulations with analytic models of biasing. We summarize our
results and conclude in Section~\ref{sec:summary}.

\section{Descriptions and Models of Biasing}
\label{sec:formalism}
A biasing scheme relates the density fluctuation fields of galaxies
and mass. For the mass, we define $\delta(\vec{x}) \equiv
[\rho(\vec{x})-\rho_{\rm b}]/\rho_{\rm b}$, where $\rho_{\rm b}$ is
the background density.  We denote the corresponding fields for the
galaxies (or halos) by $\delg(\vec{x})$. The two fields are smoothed
with the same window function of scale $\Rs$. We assume that biasing
is a local process, which means that the galaxy density field $\delg$
is related to $\delta$ within the local smoothing volume.

\subsection{Descriptions of Non-linear, Stochastic Biasing}
\label{sec:formalism:dl}
The simplest possible model for biasing is strictly linear and
deterministic:
\begin{equation}
\label{eqn:linear_bias_1}
\delg(\vec{x}) = b\,\delta(\vec{x}).
\end{equation}
A less restrictive definition of linear biasing follows from the
theory of biasing of density peaks in a Gaussian random field
\cite{kaiser:84,bbks}, which predicts that the galaxy-galaxy and
matter-matter correlation functions are related in the linear regime
by a constant multiplicative factor,
\begin{equation}
\label{eqn:linear_bias_2}
\xi_{\rm gg}(r) = b^2 \xi_{\rm mm}(r).  
\end{equation}
In both cases, $b$ is referred to as the ``linear biasing parameter''.
Eqn.~\ref{eqn:linear_bias_2} follows from Eqn.~\ref{eqn:linear_bias_1}
but the converse is not true. 

Linear deterministic bias can serve only as a crude null hypothesis:
it is not a self-consistent model, it has no theoretical motivation
and it seems inconsistent with observations. \citeN[hereafter
DL]{dekel-lahav} have proposed a general formalism for non-linear,
stochastic local biasing, which we shall adopt in this paper.  We
shall denote the one-point probability distribution functions (PDF)
for the matter and galaxy density fields as $P(\delta)$ and
$P(\delg)$.  By definition, both distributions have zero mean, and the
corresponding variances are $\sigma^2 \equiv \av{\delta^2}$ and
$\sigma_{\rm g}^2 \equiv \av{\delg^2}$.  The local biasing relation
between galaxies and matter is treated as a non-deterministic process,
specified by the conditional biasing distribution $P(\delg |
\delta)$. The mean {\it biasing function} $b(\delta)$ is then
defined by 
\begin{equation}
\label{eqn:mean_biasing_function}
b(\delta) \delta \equiv \av{\delg | \delta} = \int d\delg\, P(\delg |
\delta) \, \delg.
\end{equation}
This function fully characterizes the mean non-linear biasing and
reduces naturally to the linear biasing relation
Eqn.~\ref{eqn:linear_bias_1} if $b$ is independent of $\delta$.

Useful statistics for characterizing the mean biasing and its
non-linearity are the moments
\begin{equation}
\label{eqn:bhat}
\hat{b} \equiv \av{b(\delta) \delta^2 } /\sigma^2
\end{equation}
and
\begin{equation}
\label{eqn:bwiggle}
\tilde{b}^2 \equiv \av{ b^2(\delta) \delta^2 }/\sigma^2.
\end{equation}
To describe the statistical nature of the biasing relation, we define
the random biasing field
\begin{equation}
\label{eqn:random_biasing_field}
\epsilon = g - \av{\delg | \delta},
\end{equation}
and its variance, the biasing scatter function
\begin{equation}
\label{eqn:biasing_scatter_function}
\sigma^2_b(\delta) \equiv \av{\epsilon^2 |\delta} /\sigma^2.
\end{equation}
Averaging over $\delta$, one obtains the biasing scatter $\sigmab$:
\begin{equation}
\label{eqn:biasing_scatter}
\sigma^2_{\rm b} \equiv \langle \epsilon^2 \rangle /\sigma^2 .
\end{equation}

To second order, the three parameters $\hat{b}$, $\tilde{b}$ and
$\sigmab$ characterize any local, non-linear and stochastic biasing
relation.  The parameter $\hat{b}$ is simply the slope of the linear
regression of $\delg$ on $\delta$, and as such it is a natural
generalization of the linear biasing parameter of
Eqn.~\ref{eqn:linear_bias_1}. The ratio $\tilde{b}/\hat{b}$
quantifies the non-linearity of the mean biasing relation and
$\sigmab/\hat{b}$ independently measures the scatter. Thus the
non-linearity and stochasticity can be studied separately.  The
formalism reduces in a natural way to linear biasing when $\tilde{b} =
\hat{b}$ and to deterministic biasing when $\sigmab = 0$.  We do not
investigate moments higher than second order in this paper, but the
formalism is readily generalizable.

Various statistics describing biasing are used in the literature and
commonly referred to without distinction as ``the'' bias parameter.
In general, these statistics are not equivalent.  Here we follow the
notation of DL which naturally generalizes the linear deterministic
biasing parameter of Eqn.~\ref{eqn:linear_bias_1} into the mean
biasing parameter $\hat{b}$ of Eqn.~\ref{eqn:bhat}.  The biasing
parameter relating the variances at a given smoothing scale is denoted
$\bvar$, and it can be expressed in terms of the three basic
parameters above:
\begin{equation}
\bvar\equiv {\sigma_{g} \over \sigma} = \hat{b} \left
[\frac{\tilde{b}^2}{\hat{b}^2} + \frac{\sigma_{b}^2}{\hat{b}^2}
\right]^{1/2} \, .
\end{equation}
A complementary parameter is the linear correlation coefficient
\begin{equation}
r \equiv \frac{\langle \delg \delta \rangle}{\sigma_{\rm g} \sigma} =
\left [\frac{\tilde{b}^2}{\hat{b}^2} +
       \frac{\sigma_{b}^2}{\hat{b}^2} \right]^{-1/2} \, .
\end{equation}
Note that both $\bvar$ and $r$ mix nonlinear and stochastic
effects. We denote the ratio of correlation functions at a given
separation by $b_\xi = \sqrt{\xi_{\rm gg}/\xi_{\rm mm}}$.

\subsection{Models of Biasing}
\label{sec:formalism:models}
\subsubsection{Galaxy Conserving Models}
\label{sec:formalism:models:fry}
The bias and linear correlation coefficient are expected to evolve
with time in accord with the evolution of clustering.  If the galaxies
behave as test particles in the matter density field and their numbers
and intrinsic properties are conserved, then they satisfy the
continuity equation $\dot{\delta_{\rm g}} \simeq \dot{\delta} \simeq
-\nabla \cdot v$. The parameters $\bvar$ and $r$ then evolve as
\cite{tegmark-peebles:98}:
\begin{eqnarray}
\label{eqn:tp}
{b}_0 & = & [(1-D)^2 + 2D(1-D)b r + D^2b^2]^{1/2} \\
r_0 & = & [(1-D) + Db r]/{b}_0 \, ,
\end{eqnarray}
where $b_0$ and $r_0$ are the present day quantities, $b$ and $r$ are
the corresponding values at some earlier redshift $z$, and $D$ is the
linear growth factor at $z$ (normalized to unity today). If there is
no correlation between galaxies and mass ($r=1$), this expression
reduces to the model originally proposed by \citeN{fry:96}. We shall
refer to this type of model as ``galaxy conserving'' (GC).

\subsubsection{Hierarchical Merging Models}
\label{sec:formalism:models:mw}
\citeN[hereafter MW]{mw} developed a model for the biasing of
virialized dark-matter halos with respect to the underlying matter
distribution.  Their model is based on an extension of the
Press-Schechter approximation \cite{ps:74}, and accounts for halo
merging and continuous formation of new halos. They derived an
expression for the mean nonlinear biasing function $b(\delta; M, z,
R_s)$ of halos with mass $M$ at a redshift $z$ for a smoothing scale
$R_s$ (MW eqn. 19-20). In the vicinity of $|\delta | \gg \delta_1$,
where $\delta_1 \equiv \delta_c(1+z)$ is the density of a collapsed
object, this reduces to the simple linear biasing factor:
\begin{equation} 
\label{eqn:mw}
b(M,z) = 1 + \frac{\nu^2 -1}{\delta_1}\, ,
\end{equation} 
where $\nu \equiv \delta_1/\sigma(M)$, and $\sigma(M)$ is the rms mass
variance. MW showed that their model predictions for these quantities
are in good agreement with the results of N-body simulations with
scale-free initial power spectra, $P(k) \propto k^n$.

\subsubsection{Shot Noise and Stochasticity}
\label{sec:formalism:shot}
In empirical measures of the scatter in the biasing relation, shot
noise is inevitably mixed with the physical sources of
stochasticity. Removing this shot noise is not straightforward because
halos or galaxies correspond to rare peaks in the density
distribution, and their selection is far from a Poisson process. Let
us imagine however a deterministic biasing scheme with a known form
$\langle \delg | \delta \rangle$ (for example, measured from a
simulation). The expected variance in the number of galaxies $N$ in a
smoothing window with volume $V$ and overdensity $\delta$ is $\mu_2
\equiv \langle (N-\langle N \rangle)^2 \rangle = N=nV(1+\delg)$, where
$n$ is the global average number density of galaxies. The variance of
the galaxy field $\delg$ as a function of $\delta$ is then
$\sigma^2_g(\delta) = \mu_2/(nV)^2 = (1+\delg(\delta))/(nV)$. We shall
refer to this simple model as the ``conditional Poisson'' model.

\section{Simulations}
\label{sec:sims}
The dissipationless $N$-body simulations and the semi-analytic method
used to model galaxies within these simulations are described in
detail in \citeN{kauffmann:98a}. Here, we briefly summarize
only their main features.

\subsection{$N$-body Simulations}
\label{sec:sims:nbody}
\begin{table}
\caption{Simulation Parameters. From left to right:
the cosmological mass-density parameter $\Omega$, the Hubble constant
$h\equiv H_0/(100\, \kmsmpc)$, the linear rms mass density in a sphere
of radius $8 \hmpc$ $\sigma_8$, the box size $L$, and the mass per
particle, $m_{\rm p}$. }
\begin{center}
\begin{tabular}{lccccccc}
\hline 
Model  & $\Omega$ & $h$ & $\sigma_8$ & $L$
($\hmpc$) & $m_{\rm p}$ ($\hmsun$)\\
\hline
$\tau$CDM & 1.0 & 0.5 & 0.6 & 85 & $1.0\times10^{10}$\\
$\Lambda$CDM & 0.3 & 0.7 &0.90 & 141 & $1.4\times10^{10}$  \\
\hline
\end{tabular}
\label{tab:cosmo}
\end{center}
\end{table}

Special $N$-body simulations were run for this project (termed
``GIF'') using the version of the adaptive particle-particle
particle-mesh (AP$^3$M) Hydra code developed as part of the VIRGO
supercomputing project. The simulations have $N_{\rm p}=256^3$
particles and 512 cells on a side, and a gravitational softening
length of $30 \hkpc$ ($h \equiv H_0/(100\, {\rm km\, s^{-1}\,
Mpc^{-1}})$). Here we analyze only two of the four cosmological models
simulated for the GIF project (see Table~\ref{tab:cosmo}). One model,
\taucdm, has $\Omega=1$ and the other, \lcdm, has $\Omega=0.3$ and a
cosmological constant $\Omega_{\Lambda}=0.7$.  The initial power
spectra of fluctuations both have a shape parameter $\Gamma=0.21$ and
are normalized to approximately reproduce the number density of
clusters at the present epoch. Dark matter halos were identified using
a standard friends-of-friends algorithm with a linking length 0.2
times the mean interparticle separation, thus corresponding to a
density contrast of $\sim 125$ at the outer parts of the halos.  We do
not include any halos smaller than 10 particles in our analysis, as
tests show that these halos are not stable over many output times.

\subsection{Semi-Analytic Modelling of Galaxies}
\label{sec:sims:galaxy}
A ``merger tree'' is constructed for each halo identified at $z=0$ by
searching for its progenitor halos at earlier redshifts. A halo at an
early redshift $z_1$ is defined to be a progenitor of a halo at a
later redshift $z_0 < z_1$ if more than half of the particles of the
progenitor and its most-bound particle are included in the halo at
$z_0$. The progenitors are the halos that will merge together to form
the parent halo at $z=0$. If a halo is at the top level of the
hierarchy (i.e. it has no progenitor with at least 10 particles), then
it is assumed to contain hot gas at the virial temperature of the
halo. This gas is allowed to cool according to the radiative cooling
timescale, and the cold gas is assumed to settle into a galactic disk
and form stars.

The star formation rate is given by the expression
\begin{equation}
\label{eqn:sf}
\dot{m}_{*} = \alpha \: \frac{m_{\rm cold}}{t_{\rm dyn}},
\end{equation}
where $\alpha$ is a free parameter, $m_{\rm cold}$ is the mass of cold
gas in the disk, and $t_{\rm dyn}$ is the dynamical time of the disk.
The dynamical time is approximated as $t_{\rm dyn} = 0.1 r_{\rm vir}
/V_c$, where $r_{\rm vir}$ is the virial radius and $V_c$ is the
circular velocity of the halo at $r_{\rm vir}$. Cold gas may be
reheated by supernovae feedback, with the rate of reheating given by
the expression
\begin{equation}
\label{eqn:fb}
\dot{m}_{\rm rh} = \epsilon \left(\frac{V_0}{V_c}\right)^2 \dot{m}_{*},
\end{equation}
where $\epsilon$ is a free parameter, and $V_0$ is a scaling constant.
Reheated gas is removed from the cold gas reservoir. It may then be
retained in the halo where it will cool again on a relatively short
time scale, or ejected from the halo. If ejected, the gas is returned
to the halo after the mass of the halo has grown by a factor of two or
more.

The N-body simulations used here do not have sufficient resolution to
resolve sub-halos once they are incorporated into larger
halos. Therefore, the merging rate of galaxies within the halos is
modeled semi-analytically. When halos merge, the central galaxy of the
largest progenitor halo becomes the central galaxy of the new halo and
all other galaxies become ``satellite'' galaxies. A satellite galaxy
is assumed to merge with the central galaxy once its dynamical
friction timescale has elapsed (see \citeNP{BT}). The morphologies of
the galaxies are determined by assuming that major mergers ($m_{\rm
sat}/m_{\rm central} > 0.3)$ result in destruction of the disk and
consumption of all remaining gas in a starburst. The remnant is
assumed to be a bare spheroid. Subsequent gas cooling may result in
the formation of a new disk. The bulge-to-disk ratio at each output
redshift is then used to divide the galaxies into rough morphological
categories.

The star formation history of each galaxy is convolved with stellar
population synthesis models to obtain total luminosities in any
desired filter band. The models of \citeN{gissel:98} are used,
assuming that all stars have solar metallicity, and a standard Scalo
initial mass function \cite{scalo:86}. The effects of dust extinction
on the galaxy spectra were investigated using the empirical recipe of
\citeN{wang-heckman} (WH), in which the face-on optical depth in the
B-band is parameterized in the form:
\begin{equation}
\label{eqn:wh}
\tau_B \propto \tau^{*}_B (L_B/L_B^{*})^\beta\, , 
\end{equation}
with $\beta=0.5$, $\tau^{*}_B=0.8$, and $L_B = 1.3 \times 10^{10}
h^{-2} L_{\odot}$. This can be extended to other wavebands using a
standard Galactic extinction curve. The extinction is then computed by
assigning a random inclination to each galaxy and using a standard
``slab'' model (see \citeN{kauffmann:98a} for details).

The two free parameters in the galaxy formation models ($\alpha$ in
Eqn.~\ref{eqn:sf} and $\epsilon$ in Eqn.~\ref{eqn:fb}) are set by
requiring the central galaxy in a halo with $V_c = 220\, \kms$ at the
present epoch to have an I-band magnitude of $M_{I} -5\log h \sim
-22.1$ and a cold gas mass of $m_{\rm cold}\sim 10^{10} \msun$. This
ensures that the zero-point of the Tully-Fisher relation for spiral
galaxies is in agreement with observations
(e.g. \citeNP{giovanelli:97}).

Various properties of the galaxies produced using these techniques are
summarized in the companion papers based on the GIF simulations,
\citeN{kauffmann:98a}, \citeN{kauffmann:98b}, and \citeN{diaferio:98}.
We note that throughout this paper, we include in our analysis only
galaxies that reside in halos of least 10 particles.

\section{From Halos to Galaxies}
\label{sec:masstolight}
The study of biasing is in some sense the attempt to understand the
connection between the total mass density, which we believe to be
dominated by dark matter, and the density of luminous galaxies. A
fairly robust component of modern galaxy formation theory is that
galaxies form within collapsed, virialized dark-matter halos. The
clustering properties of these halos relative to the underlying mass
distribution are straightforward to compute from cosmological $N$-body
simulations such as the ones that we have described, subject only to
the numerical limitations of the simulation at hand. Predicting how
these halos are connected to visible \emph{galaxies} is far more
difficult. We have adopted the particular recipes described above, but
other equally plausible recipes could lead to different results. In
this section we quantify the resulting statistical relation between
halos and galaxies in our simulations, and in the next section we
present in parallel the results for the biasing properties of halos
and galaxies. In this way we hope to gain some insight into the extent
to which the biasing of galaxies is determined by the gravitational
process of halo formation and to what extent by other astrophysical
processes.

In general, a single dark-matter halo in our simulations may contain
several galaxies with a range of luminosities.  The largest collapsed
halos in our simulations have virial radii of $1.5-2
\hmpc$. Therefore, on the smoothing scales that investigate here ($R_s
\ga 4 \hmpc$), the clustering of galaxies is mainly determined by the
clustering of the halos in which they
dwell. Figure~\ref{fig:massmag_bdust} shows the joint probability
function of absolute magnitude and host halo mass for the galaxies in
our simulations. The pronounced diagonal ``ridge'' is populated by
galaxies that are the central object in their halo, and illustrates
the fairly tight relationship between luminosity and halo mass or
circular velocity (i.e. the Tully-Fisher relation) obeyed by these
galaxies. This ridge corresponds to a constant halo mass-to-light
ratio, where the value of the constant is constrained to agree with
the zero-point of the observed Tully-Fisher relation by construction
(see Section~\ref{sec:sims:galaxy}). The galaxies that lie far off of
the ridge are mainly satellites. Thus we see that galaxies with a
given luminosity live in a wide range of environments. The differing
appearance of the contours in the two different cosmologies (for
example the much broader ridge in the \taucdm\ simulations) are a
direct consequence of the details of the star formation and supernovae
feedback recipes, which were chosen to give the best agreement with
observations \cite{kauffmann:98a}. Put another way, these two
cosmologies have different halo mass functions, so that in order for
both to reproduce the observed galaxy luminosity function, the
relationship between halos and galaxiesmust be different. The
right-hand panels show the results of including dust extinction. We
see that this effect can substantially modify the relationship between
mass and light in the B-band. We return to this point later.

Figure~\ref{fig:mlhalos} shows the mass-to-light ratio of halos in our
simulations in the B and I bands. The effective mass-to-light ratio
when dust extinction is included is also shown. The mass-to-light
ratio varies by more than one order of magnitude from the smallest to
largest halos in the simulations, and has a large scatter at a fixed
halo mass. The characteristic mass $M_{*}$, defined as the mass that
is just becoming non-linear at a particular epoch (more precisely
$\sigma(M_{*}(z)) = \delta_c(z)$), is $M_{*} = 2.9 \times 10^{12}
\hmsun$ for the \taucdm\ model and $M_{*} = 1.5 \times 10^{13} \hmsun$
for the \lcdm\ model (see figure~\ref{fig:mstar}).  The exponential
turn-over in the halo mass function thus occurs at a mass much larger
than that of halos that typically host $L_*$ galaxies. Therefore the
sharp increase in $M/L$ at $\sim10^{13} \hmsun$ is necessary in order
to produce a luminosity function with a ``knee'' at $L_{*}$.  This is
generally believed to be the result of inefficient cooling in large
halos, and in these simulations it is achieved by turning off gas
cooling in halos larger than 350 \kms (see \citeNP{kauffmann:98a}). We
may also note that the difference between the mean B-band and I-band
mass-to-light ratio increases with halo mass, indicating that larger
halos tend to host galaxies with redder colours.

The distribution $P(n | M_h)$ of the number of bright galaxies per
halo as a function of halo mass is shown in figure~\ref{fig:ngalmass}.
Halos with masses between $10^{12} \hmsun$ and $10^{13} \hmsun$
contain one bright galaxy on average, with a relatively small
dispersion. Larger mass halos ($\ga 5\times 10^{13} \hmsun$) typically
contain more than one galaxy per halo, and the number of galaxies per
halo has a larger dispersion; these halos represent groups or clusters
of galaxies.

\section{Biasing of Halos and Galaxies} 
\label{sec:results}
In this section we present our main results concerning the biasing
relation between the overdensity fields of halos, or galaxies, and
that of the underlying dark matter.

Figure~\ref{fig:dcontourh} shows the joint distribution $P(\delg,
\delta)$ for halos, and figure~\ref{fig:dcontourg} shows the same for
galaxies. All the density fields have been smoothed with a top-hat
filter of radius 8 \hmpc\ (hereafter T8).  We have selected halos with
masses larger than $1.0\times10^{12} \hmsun$, and galaxies with $M_B-5
\log h \le -19.5$, at several redshifts as indicated on the figures.
The conditional mean and standard deviation functions are shown,
corresponding to $b(\delta)$ and $\sigmab(\delta)$ of
Eqn.~\ref{eqn:mean_biasing_function}
and~\ref{eqn:biasing_scatter_function}.  Shown for reference is also
the linear biasing approximation $\delg = b \delta$, with $b=\bhat$.
A log-log density plot is used in order to stress the behavior in the
regions of under-density.  Note that in this specific presentation,
linear biasing appears as a curved line.

The results for halos and galaxies are qualitatively similar. This is
somewhat surprising given the rather complex relationship between the
two that we saw in the previous section. However, recall from
figure~\ref{fig:ngalmass} that \emph{on average} there is
approximately one galaxy brighter than $-19.5+5 \log h$ per $\sim
10^{12} \hmsun$ halo. These halos are much more numerous than the
larger mass halos that host larger numbers of galaxies, so they tend
to dominate the appearance of a number-weighted joint probability such
as the quantity represented in these figures.  We also note that the
contours for the two cosmological models differ in their details but
overall they appear qualitatively similar.

The favored environment for halo/galaxy formation, namely, the peak in
the halo/galaxy overdensity distribution, occurs in regions where the
matter density is close to its mean value ($1+\delta=1$).  In the
vicinity of $1+\delta \sim 1$, the overdensity of halos/galaxies
follows that of the underlying matter, $b(\delta)\simeq 1$. But, in
general, the linear biasing approximation is not an accurate
description of the mean biasing relation.

The mean biasing function shows a robust characteristic behavior in
the under-dense regions ($1+\delta<1$); the local slope of $b(\delta)$
is steeper than the global slope $\hat{b}$, and is larger than unity
even when $\hat{b}<1$.  This leads to voids empty of galaxies below
some finite matter underdensity ($1+\delta>0$); namely, the
galaxy-formation efficiency drops to zero below a certain mass-density
threshold. We can say that the galaxies in the voids are always
``positively biased'', in the sense that locally $\hat{b}>1$, while
they can be either biased or ``anti-biased'' ($\hat{b}<1$) in the
high-density regions.

In the over-dense regions ($1+\delta>1$), the behavior changes with
redshift. At $z=0$, the local slope of $b(\delta)$ is smaller than
unity, driving $\hat{b}$ to below unity.  At higher redshifts the
local slope becomes larger, driving $\hat{b}$ to values as large as 3
to 7 by $z=3$.  The redshift dependence of the biasing relation is
striking; we return to it in \ref{sec:results:redshift}.

The stochasticity in the biasing scheme is evident from the 
spread in $\delg$ at fixed $\delta$.  We will discuss it further in
the next section.

\subsection{Mass and Luminosity Dependence} 
\label{sec:results:masslum} 
In this section we investigate the dependence of biasing on halo mass
and galaxy luminosity, for a fixed smoothing window (T8) and at the
present epoch ($z=0$).  Figure~\ref{fig:moments_mass} (bottom) shows
the mean biasing function for halos selected with different mass
thresholds, in the \taucdm\ simulation.  We see that massive halos are
more biased; the curves all cross at $\delta\simeq 0$, massive halos
have higher overdensities than smaller mass halos in regions with
$\delta>0$, and smaller mass halos have higher overdensities where
$\delta<0$.

The top panel shows the variance of the conditional biasing
distribution $\av{\epsilon^2 |\delta}$ minus the mean shot noise
correction $1/(\bar{n}V_{s})$, where $\bar{n}=N/V_{\rm BOX}$ is the
average number density of halos/galaxies above the given threshold in
the simulation, and $V_s$ is the volume of the smoothing window. The
conditional Poisson model discussed in
Section~\ref{sec:formalism:shot} is shown for comparison.  In regions
of lower than average overdensity, the scatter is generally smaller
than the mean shot noise, and in overdense regions it is larger.

This can be understood by referring to the expression for the variance
of counts in cells \cite{peebles:80}:
\begin{equation}
\mu_2 \equiv \langle (N-nV)^2\rangle = nV + n^2 \int_{V} dV_1 dV_2 .
\xi_{12}
\end{equation}
The variance of the conditional biasing distribution is then
$\mu_2/(nV)^2$.  This reduces to the usual shot noise value
$1/\sqrt{nV}$ in the absence of correlations ($\xi = 0$), but it leads
to a scatter larger than the mean shot noise when the integral over
$\xi$ is positive, and smaller than the mean shot noise when it is
negative. In underdense regions ($\delta < 0$), halos are
anti-correlated ($\xi < 0$) and the volume-averaged correlation
function, $\bar{\xi}$, is negative. In overdense regions ($\delta >
0$), halos are positively correlated and $\bar{\xi}$ is positive.

The corresponding results for the \lcdm\ model are extremely similar (and are
therefore not shown) because the mass dependence enters only through the shape
of the power spectrum, which is similar by design for the \taucdm\ and \lcdm\
simulations.

Figure~\ref{fig:moments_lum} summarizes the mean and scatter of the
conditional biasing relation for \emph{galaxies} with different
luminosity thresholds. Two notable features are evident. First, we see
no significant change in the biasing relation for galaxies with
absolute-magnitude thresholds in the range $-18.4+5\log h$ (well below
$L_{*}$) to $-19.9 +5\log h$ (above $L_{*}$). We return to this point
in a moment, when we discuss figure~\ref{fig:xig_lum}.  Second, the
scatter compared to the average shot noise is comparable to the
corresponding quantity for halos. On the face of it, one might think
that the complex physics of gas cooling, star formation, merging,
supernovae feedback, etc, even as simply modelled in our simulations,
would lead to a larger scatter. However, on the other hand, the nature
of the physics associated with these processes might actually lead to
a stronger correlation of luminosity with the local matter density
than that shown by the halos. For example, gas can only cool and form
stars in regions with sufficiently high density, the star formation
efficiency is assumed to be proportional to the density, supernovae
feedback is less efficient in halos with deep potential wells
(i.e. high density), and starbursts occur preferentially in high
density regions. While these lines of argument appear plausible, more
detailed investigation of the physical source of scatter in the
biasing relation that we obtain both for galaxies and halos is clearly
in order.

The three characteristic parameters of the biasing scheme (see
Section~\ref{sec:formalism:dl}), measuring mean biasing, nonlinearity,
and stochasticity, are shown in figure~\ref{fig:bias_masslum} as a
function of halo-mass threshold or galaxy luminosity threshold.  The
mean biasing is characterized by $\hat b$, and also shown are $\bvar$
and $\bxi$.  The fact that $\bvar$ tends to be systematically larger
is due to the presence of nonlinearity and stochasticity.  The
nonlinearity, as measured by $\tilde{b} /\hat b$, is less than 10
percent for halos and is nearly constant over the mass range. The
nonlinearity is even smaller for galaxies, and shows a trend with
galaxy luminosity. The stochasticity, $\sigma_{\rm b}/\hat b$, shows a
modest trend, increasing for higher mass halos or brighter
galaxies. As remarked before, the stochasticity is actually larger for
halos than for galaxies. The linear correlation coefficient $r =
\bhat/\bvar$, which, like $\bvar$, mixes nonlinear and stochastic
effects, is roughly constant over these ranges of mass or luminosity
and has a value $r \simeq 0.9$.

We further examine the clustering of halos of different masses
compared to the dark matter in figure~\ref{fig:xi_mass}, which shows
the auto-correlation functions of halos and dark matter in the
\taucdm\ simulations. More massive halos show correlations of higher
amplitude and therefore more positive biasing. The turn-over of the
halo correlation functions at small radii is caused by exclusion
effects due to the finite spatial extent of the halos. When using
conventional halo finders based on friends-of-friends or spherical
over-density, halos that lie too close together are ``merged'' into a
single halo. The spatial scale at which this effect becomes important
depends on the mass of the halo, according to the relation between
virial radius and mass ($M_{\rm vir} \propto r^3_{\rm vir}$). Thus on
small scales, one expects anti-correlations due to this exclusion
effect to lead to negative $\bar{\xi}$ in regions of high halo
density, and thus to a reduction in the scatter, but this effect is
negligible for the smoothing scales considered here ($\Rs \ga 4
\hmpc$)\footnote{The exclusion effect results in a reduction of the
scatter by a factor that scales like $(r_{\rm ex}/\Rs)^3$, where
$r_{\rm ex}$ is the scale on which the exclusion effects are
important. Here $(r_{\rm ex}/\Rs)^3 \simeq 2 \times 10^{-2}-2\times
10^{-3}$ depending on the halo mass, for $\Rs = 4 \hmpc$.}. Note that
the fact that the observed galaxy correlation function is a nearly
perfect unbroken power-law on sub-Mpc scales indicates that there must
be a significant contribution from galaxy pairs that dwell within a
common halo.

The correlation function of galaxies with different luminosity
thresholds is shown in figure~\ref{fig:xig_lum}\footnote{The galaxy
catalogs used here are not identical to the ones used in
\citeN{kauffmann:98a}, which explains the small differences in the
correlation functions shown here and Figure~10 and 11 of
\citeN{kauffmann:98a}.}. As noted by \citeN{kauffmann:98a}, the
clustering amplitude of the galaxies in the simulations is nearly
independent of the luminosity threshold, in contrast to the findings
of \citeN{willmer:98}, who observed pronounced luminosity-dependent
bias in the SSRS2 redshift survey. They found that brighter galaxies
are more strongly clustered than fainter galaxies, and thus have a
longer correlation length $r_0$ as shown on the figure.

\begin{figure*}
\centerline{\psfig{file=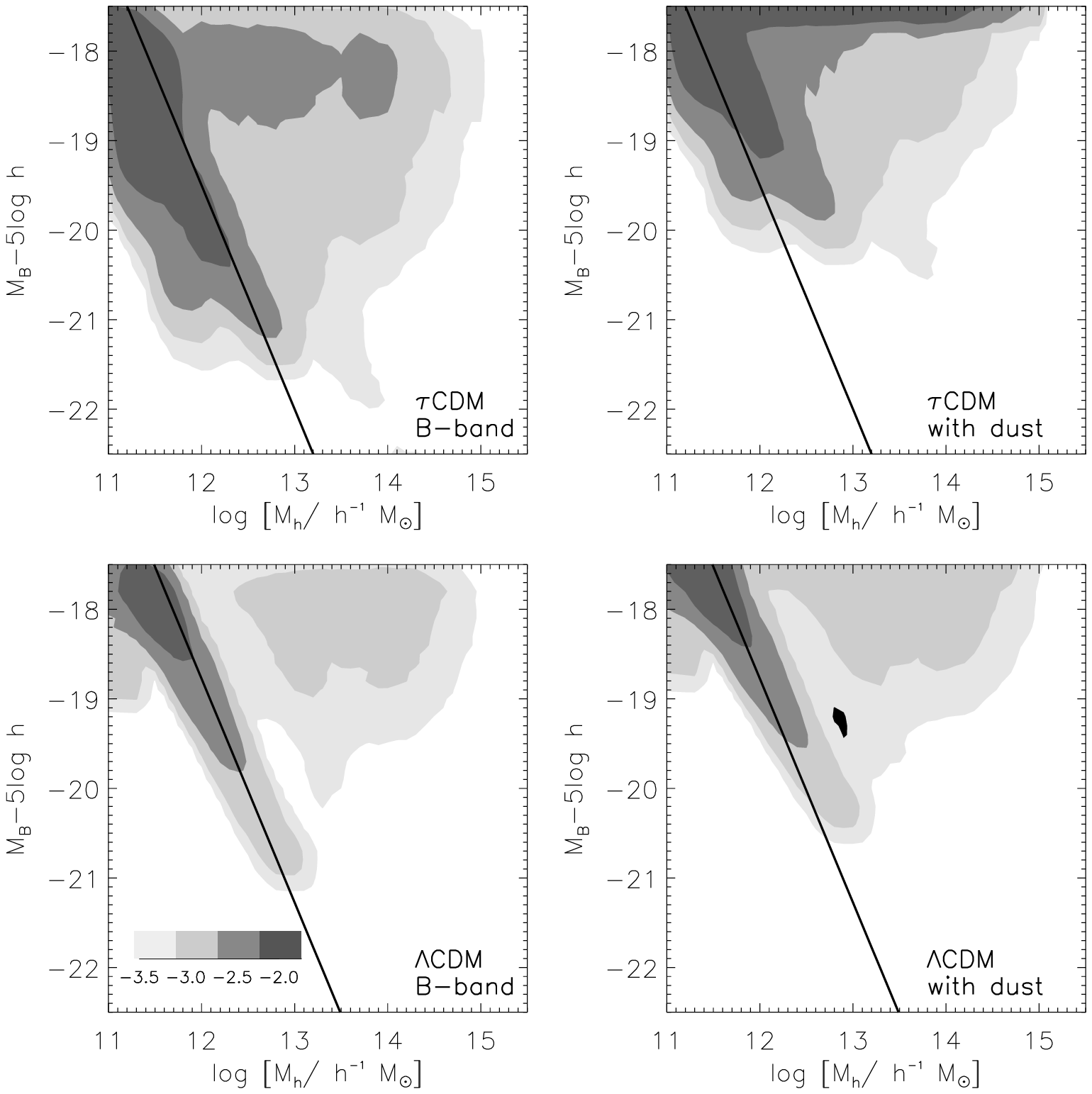,height=16truecm,width=16truecm}}
\caption{The relationship between galaxy luminosity and the mass of the host
halo. The contours represent the joint probability that a galaxy has a given
luminosity and dwells in a host halo with a given mass. The grey-scale
indicates the $\log_{10}$ of the probability, as shown by the scale bar on the
figure. The right panels show the effects of applying a differential correction
for dust extinction to the modelled luminosities, in which more luminous
galaxies are more extinguished (see text). Top panels show the \taucdm\
simulations and bottom panels show the \lcdm\ simulations. The diagonal lines
indicate constant mass-to-light ratio, with the constant dictated by the
observed B-band Tully-Fisher relation.}
\label{fig:massmag_bdust}
\end{figure*}

\begin{figure*}
\centerline{\psfig{file=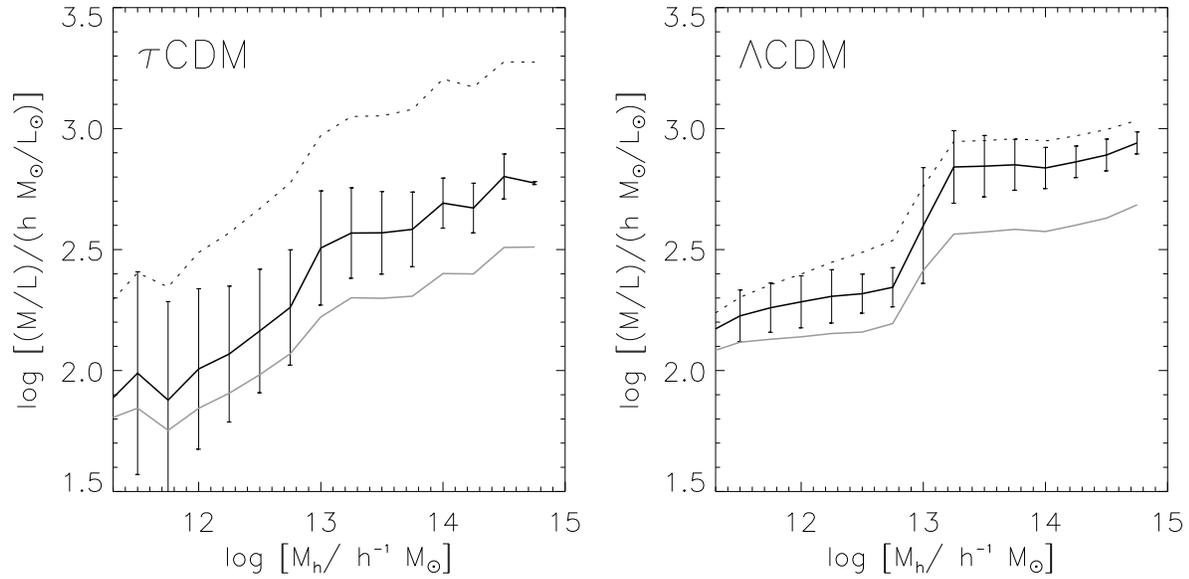,height=8truecm,width=16truecm}}
\caption{The mass to light ratio of halos in the simulations. The solid black
line shows the B-band $M/L$, and error bars show the 1-sigma scatter. The
dotted line shows the B-band $M/L$ after a correction for dust extinction has
been applied (see text). The grey line shows the I-band $M/L$ (without dust).
}
\label{fig:mlhalos}
\end{figure*}

\begin{figure}
\centerline{\psfig{file=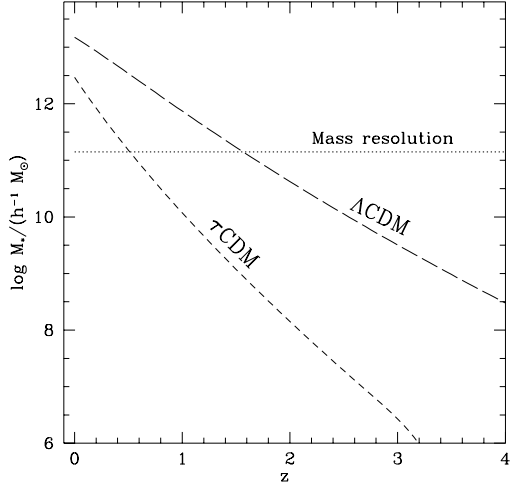,height=7truecm,width=7truecm}}
\caption{The characteristic non-linear mass $M_*$ (see text) as a 
function of redshift for the \taucdm\ and \lcdm\ models. The horizontal
dotted line indicates the mass of the smallest halos that we can
reliably resolve in our simulations. }
\label{fig:mstar}
\end{figure}

\begin{figure}
\centerline{
\psfig{file=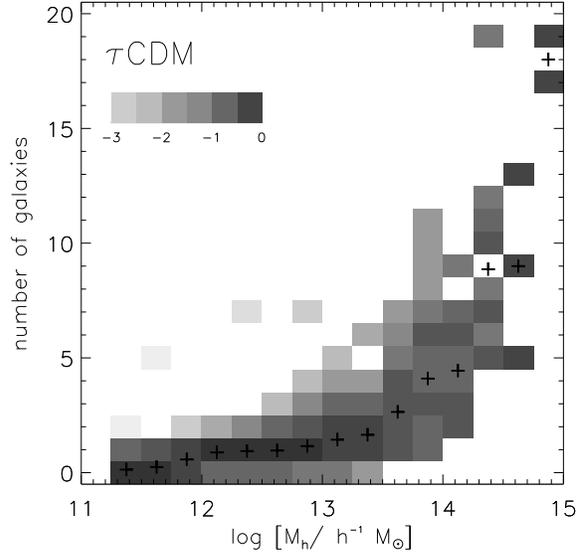,height=8truecm,width=8truecm}}
\centerline{
\psfig{file=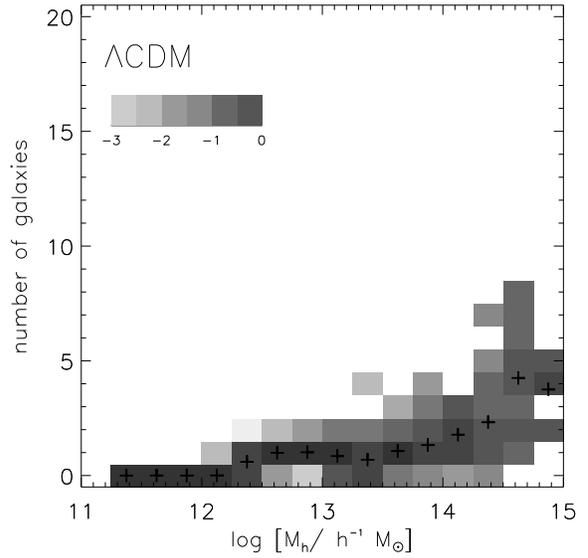,height=8truecm,width=8truecm}}
\caption{The probability of finding $n$ galaxies per halo, with 
$M_B-5\log h \le -19.5$, as a function of halo mass, in the \taucdm\
(top) and \lcdm\ (bottom) simulations. The grey scale indicates the
$\log_{10}$ of the probability, according to the scale shown on the
figure. The mean is shown by the cross symbols. }
\label{fig:ngalmass}
\end{figure}

Our result is not surprising given the weak mass dependence of the
biasing of galaxy-mass halos (figure~\ref{fig:bias_masslum}). However,
this introduces an apparent problem when compared to observations.
The difference between the faintest and brightest magnitude thresholds
for the observed galaxy samples analyzed by \citeN{willmer:98} is
about 1.5 magnitudes. This corresponds to a factor of four in mass if
the mass-to-light ratio is constant and its value is dictated by the
observed Tully-Fisher relation.  On the other hand, we saw from
figure~\ref{fig:bias_masslum} that there is only a $\sim 10\%$
increase in the mean biasing as a function of halo mass over this
range, and even this weak mass dependence is diluted by the presence
of low-luminosity satellite galaxies in large-mass halos (see
figure~\ref{fig:massmag_bdust}), yielding no luminosity dependence in
our simulated galaxy biasing.

A clue for a resolution of this problem may come from the fact that
the mass dependence of halo biasing becomes much stronger in the
regime where $M > M_{*}$ (figure~\ref{fig:bias_masslum}).  If we
increase the mass-to-light ratio, pushing the bright galaxies into
larger mass halos, we would obtain stronger luminosity dependence, as
was in fact found by \citeN{kns} when they effectively normalized
their models with a higher mass-to-light ratio.  Unfortunately, this
would lead to an apparent contradiction with the observed Tully-Fisher
relation.

A more promising way to achieve the desired effect is by inclusion of
dust in our considerations. We note that the observed Tully-Fisher
relation has been corrected for dust extinction using observationally
determined inclinations for each galaxy. These corrections can be
quite large in the B-band, where most of the samples used to study
galaxy clustering are selected. Moreover, there is evidence that more
luminous galaxies suffer larger extinctions
\cite{tully:99,wang-heckman}. Studies of galaxy clustering do not
include corrections for dust extinction because inclination estimates
are generally not available. \citeN{kauffmann:98a} showed that
including dust extinction can significantly modify the galaxy-galaxy
correlation function found in these simulations.

The \taucdm\ model suffered from an excess of bright galaxies compared to the
observed B-band luminosity function when dust extinction was
neglected. \citeN{kauffmann:98a} found that including dust extinction using the
empirical recipe of \citeN{wang-heckman} led to an improved luminosity
function, but still with an excess, especially on the bright end. We now tune
the parameters of the WH recipe in order to obtain the best possible fit to the
observed luminosity function. For the \taucdm\ model, we find that using
$\beta=0.2$ and setting $\tau^*_B = 2.0$ in eqn.~\ref{eqn:wh} gives the best
results. For the \lcdm\ model, the luminosity function already has a small
deficit of galaxies compared to observations even without any dust
correction. Despite this, we show the results of applying a dust correction to
the \lcdm\ models using the fiducial values of the WH parameters (as in
\citeNP{kauffmann:98a}).

The inclusion of dust extinction changes the effective mass-to-light
ratio of the halos in the simulations (see
figure~\ref{fig:massmag_bdust}), in the desired sense: galaxies of a
given observed luminosity now dwell in larger mass halos. We see in
the right-hand panels of figure~\ref{fig:xig_lum} that this leads to a
luminosity dependence that is qualitatively similar to the observed
dependence, especially in the $\tau$CDM case (unfortunately, the
number of bright galaxies in our simulation box also becomes quite
small, and the correlation function becomes rather noisy, but the
trend is clear). Note also that although the inclusion of dust
increases the correlation amplitude of the galaxies in the \taucdm\
model, bringing the results into better agreement with the observed
galaxy correlation function, the clustering amplitude for bright
galaxies is still not as high as the SSRS2 observations. Conversely,
with no dust correction, the clustering amplitude of galaxies for the
\lcdm\ model is already comparable to the observational results, and
inclusion of dust hardly makes a difference.  It may be that a
cosmology with an intermediate value of $\Omega_0\simeq 0.5$ will give
the best overall results when realistic extinction due to dust is
included.  However, a much larger correction seems to be required in
order to match the correlation function of very bright galaxies as
provided by these observations.  These results highlight the
importance of obtaining large redshift surveys selected in near IR
bands in order to reduce the sensitivity to dust extinction.

\subsection{Type Dependence}
\label{sec:results:type}
We study the type dependence of biasing with T8 smoothing, for
galaxies brighter than $M_B - 5\log h = -18.4$, at $z=0$.  Here,
instead of quantifying the bias of galaxies with respect to mass as in
the previous section, we show the biasing of different types of
galaxies relative to the overall galaxy population. This
\emph{relative} bias is of particular practical interest because it
can be compared directly with the results of observations.

We can classify different types of galaxies in our simulations in
several ways.  For example, we identify ``early'' and ``late'' types
according to the bulge-to-total luminosity ratio; galaxies of $L_{\rm
bulge}/L_{\rm tot} > 0.4$ are identified with early type galaxies
(Hubble type E--S0), and the rest with late types (S--Irr), as in
\citeN{kauffmann:98a}. Similarly, we can divide the galaxies in terms
of colour. Here, we classify galaxies with $B-V > 0.8$ as ``red'' and
the remainder as ``blue''. Figure~\ref{fig:relbias} (top) shows the
joint density distribution for galaxies of early and late types
relative to the galaxy population as a whole at the same magnitude
limit. Early type galaxies are biased ($\hat{b}>1$) compared to the
overall population, wheras late type galaxies are slightly anti-biased
($\hat{b}<1$) We obtain comparable results, with stronger bias and
anti-bias, when galaxies are divided in terms of their colors, as
shown in the bottom panels. The results shown are for the \taucdm\
simulation, and are similar for the \lcdm\ simulation.

Once again, we can understand this result in terms of the masses of
the dark matter halos in which these different types of galaxies are
found. Figure~\ref{fig:masshist} shows the distributions of host halo
masses for early and late type galaxies and red and blue
galaxies. Although each type occupies a broad range of halo masses,
the mean halo mass occupied by early/red types is significantly larger
than that occupied by late/blue types. Several physical effects
included in the semi-analytic modelling may contribute to this
result. Larger mass halos are more likely to have suffered a recent
major merger, which is assumed to destroy the disk and create a
bulge. In addition, gas cooling ceases in large mass halos, cutting
off the supply of new gas and subsequently the star
formation. Therefore the galaxies in these halos redden, are not able
to form new disks and so remain bulge-dominated.  Note that the
separation between the distributions of red and blue galaxies is
larger than the separation between early and late types
(figure~\ref{fig:masshist}), explaining why the relative biasing is
stronger for the former (figure~\ref{fig:relbias}).

The mean relative biasing parameters found in the simulations are
summarized in Table~\ref{tab:relbias}. These parameters are the
equivalent of $\bhat$ and $\bvar$ as defined in
Section~\ref{sec:formalism:dl}, where the matter field $\delta$ is
replaced with the field of late/blue type galaxies and $\delg$ is
replaced with the field of early/red type galaxies. The difference
between $\bhat^{\rm rel}$ and $\bvar^{\rm rel}$ again indicates the
presence of nonlinearity and stochasticity in the relative biasing
relation, and is also reflected in the relative linear covariance
parameter $r^{\rm rel}$.
\begin{table}
\caption{Relative bias of early to late and red to blue galaxies in the
simulations on a scale of 8 \hkpc. For each parameter, the first value is for
\taucdm\ and the second is for \lcdm.}
\begin{center}
\begin{tabular}{lcccccc}
\hline
type & \multicolumn{2}{c}{$\bhat^{\rm rel}$}
 & \multicolumn{2}{c}{$\bvar^{\rm rel}$}
 & \multicolumn{2}{c}{$r^{\rm rel}$}\\  
\hline
early/late & 1.4 & 1.1 & 1.5 & 1.3 & 0.93 & 0.87\\ 
red/blue & 1.6 & 1.6 & 1.8 & 1.8 & 0.87 & 0.87 \\
\hline
\end{tabular}
\label{tab:relbias}
\end{center}
\end{table}

Observationally, the stronger clustering of early type galaxies is
well known
\cite{dressler:80,hermit:96,willmer:98,tegmark-bromley:98}. In order
to compare with theoretical predictions, care must be taken to account
properly for the transformation from redshift space to real space, as
early and late type galaxies are known to be affected differently by
redshift distortions. In real space, \citeN{willmer:98} find $b_{\rm
early}/b_{\rm late}=1.18 \pm 0.15$, \citeN{loveday:95} find $b_{\rm
early}/b_{\rm late}=1.33$ and \citeN{guzzo:97} find $b_{\rm
early}/b_{\rm late}=1.68$. Not surprisingly, the range of
observational values seems to depend on the importance of clusters in
the sample, with higher values obtained in samples that include many
rich clusters. The quoted observational statistics correspond to
$\bvar$. The values we obtain (1.3-1.5) are in reasonable agreement
with the observational values.

It is also fairly well established that red galaxies are generally
more clustered than the blue population \cite{landy:96,tucker:97}, but
a quantitative comparison is more difficult, because most of the
values in the literature are obtained from the angular or redshift
space correlation functions, and a variety of colour bands are
used. \citeN{willmer:98} quote a relative variance bias in real space
$b_{\rm red}/b_{\rm blue}=1.40 \pm 0.33$ at $8 \hmpc$, and suggest
that there is evidence for scale dependence in the relative
bias. However, they used a colour threshold of $B-R=1.3$ so it is not
directly comparable with our results.

In principle, the type dependence could be further studied as a
function of redshift, scale, luminosity, or environment. The samples
currently available from both simulations and observations are too
small to obtain proper statistics after this sort of subdivision, but
this may be a powerful discriminatory tool for galaxy formation models
once larger simulations and larger redshift surveys are
available. Another promising approach would be to categorize galaxies
using automatic spectral classification methods such as Principal
Component Analysis \cite{connolly:95,folkes:96} and study the
corresponding relative bias. An example of a comparison of this sort
appears in \citeN{tegmark-bromley:98}.

\subsection{Scale Dependence} 
\label{sec:results:scale}
Figure~\ref{fig:moments_scale_h} and~\ref{fig:moments_scale_g} show
the scale dependence of biasing for a fixed halo mass threshold ($M
\ge 10^{12} \hmsun$) or galaxy luminosity ($M_B-5 \log h \le -19.5$)
at the present epoch ($z=0$).  The results shown are for top-hat
smoothings of 4, 8 and 12$\hmpc$. This figure shows the mean and
scatter of the conditional biasing relation for halos in the \taucdm\
simulation (again, the results of the \lcdm\ simulation are nearly
identical and are not shown).  The scale dependence of the mean
biasing is weak over this range, while the scatter shows an expected
scale dependence.  Once again, the results for galaxies are similar to
the results for halos. This suggests that, at least in our
simulations, the scale dependence (or lack thereof) of biasing of
bright galaxies is mainly determined by the gravitational physics of
halo formation.

Figure~\ref{fig:bias_scale} presents the scale dependence of the
biasing parameters.  Here we see that the mean biasing actually
increases slightly with smoothing scale for both halos and
galaxies. As expected, the non-linearity and stochasticity both
decrease with increasing smoothing scale, such that the biasing
relation converges towards linear deterministic biasing for large
smoothing scales.

\subsection{Redshift Dependence} 
\label{sec:results:redshift}
\begin{figure*}
\centerline{
\psfig{file=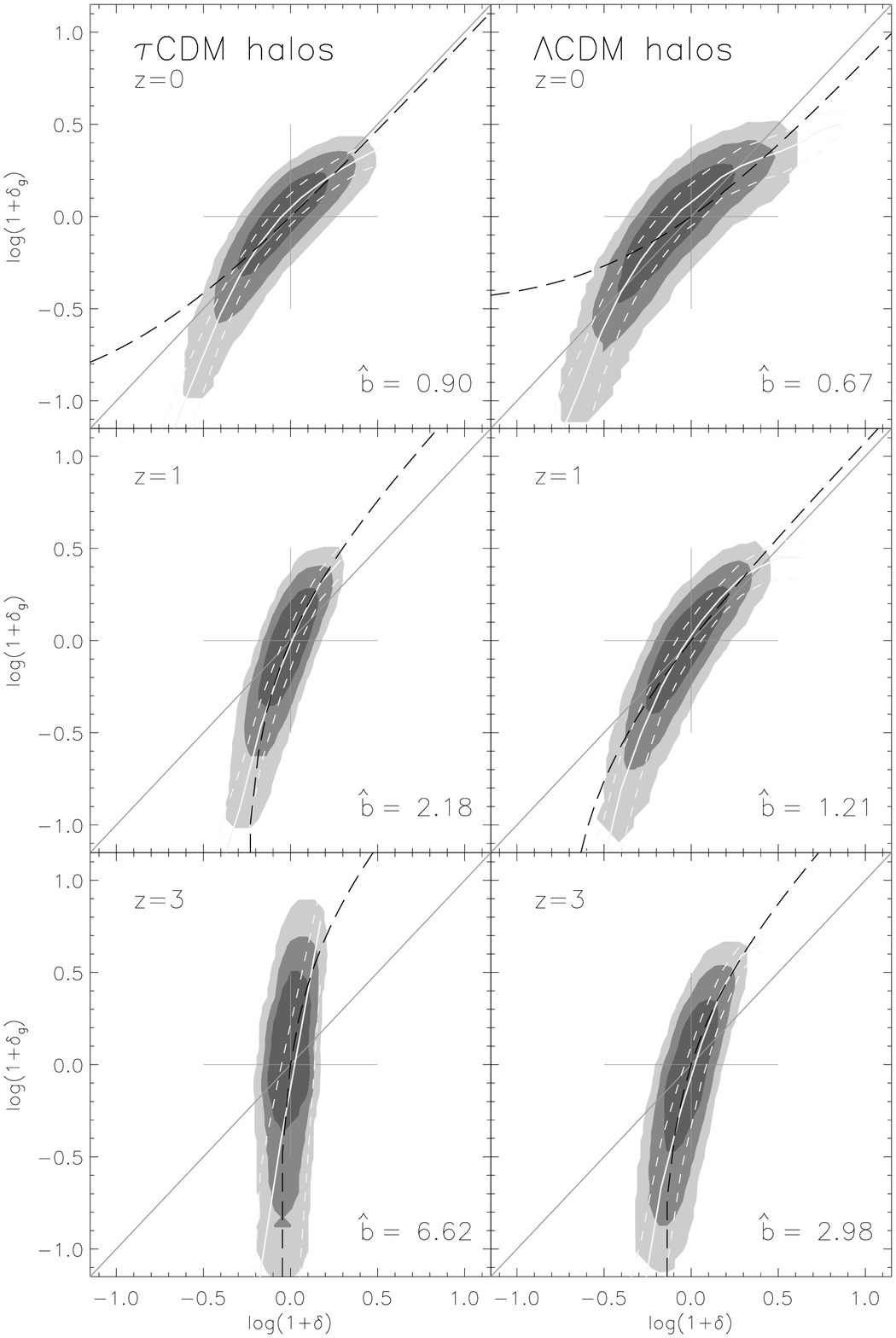,height=22truecm,width=15truecm}}
\caption{The joint distribution of the overdensity fields of halos ($M
\ge 10^{12} \hmsun$) and mass, both smoothed with a top-hat window of
radius $\Rs=8 \hmpc$, for \taucdm\ (left) and \lcdm\ (right), for
several redshifts. The contours represent approximately the 50 (dark
grey), 80 (medium grey), and 98 (light grey) percentiles. The white
lines show the mean conditional biasing function $\av{\delg | \delta}$
and the $1\sigma$ scatter about it in equal bins of
$\log(1+\delta)$. The black long-dashed line shows the linear biasing
approximation with $b=\bhat$. }
\label{fig:dcontourh}
\end{figure*}

\begin{figure*}
\centerline{\psfig{file=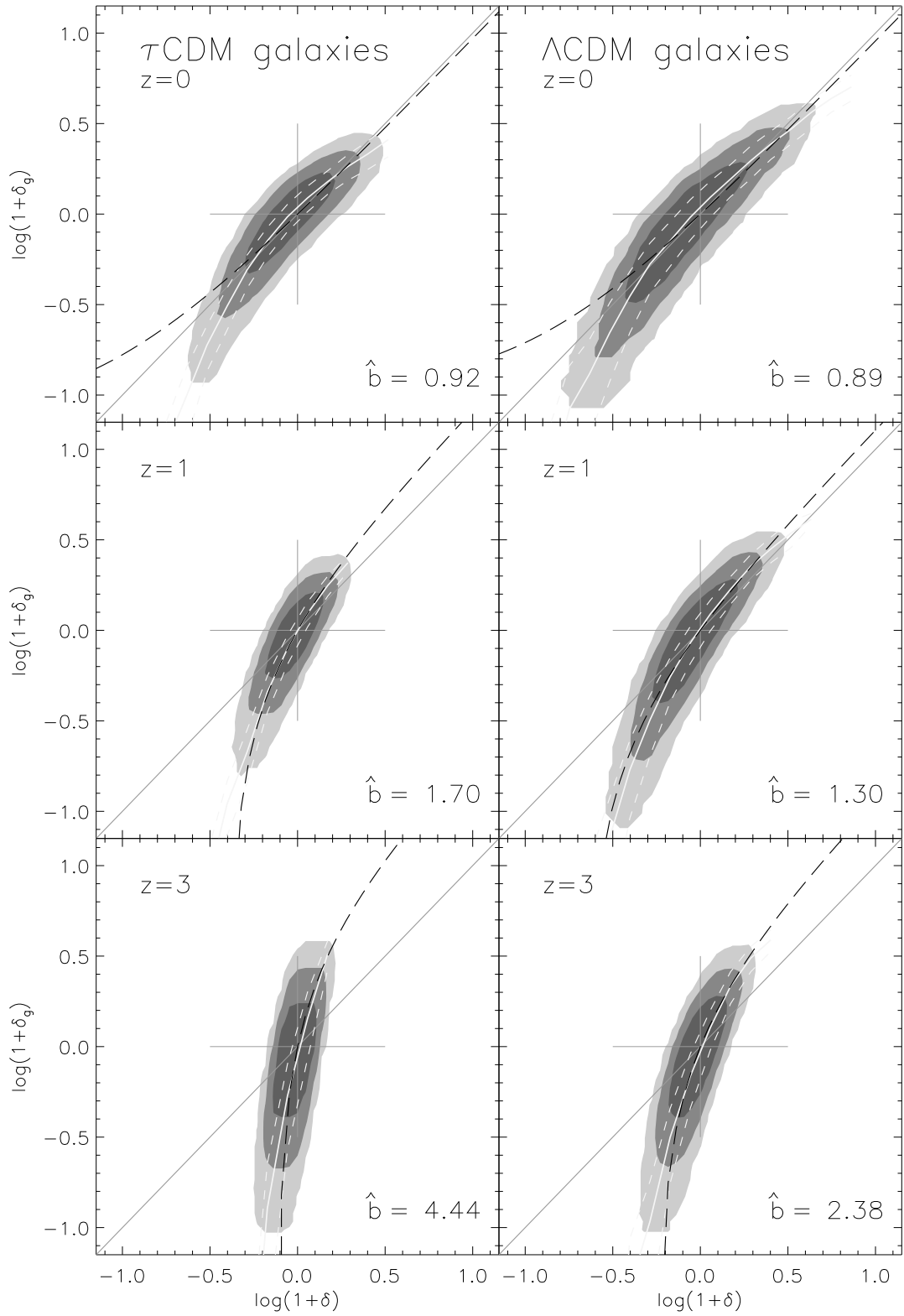,height=22truecm,width=15truecm}}
\caption{Same as figure~\protect\ref{fig:dcontourh}, 
for galaxies brighter than $M_B-5\log h = -19.5$. }
\label{fig:dcontourg}
\end{figure*}

\begin{figure}
\centerline{
\psfig{file=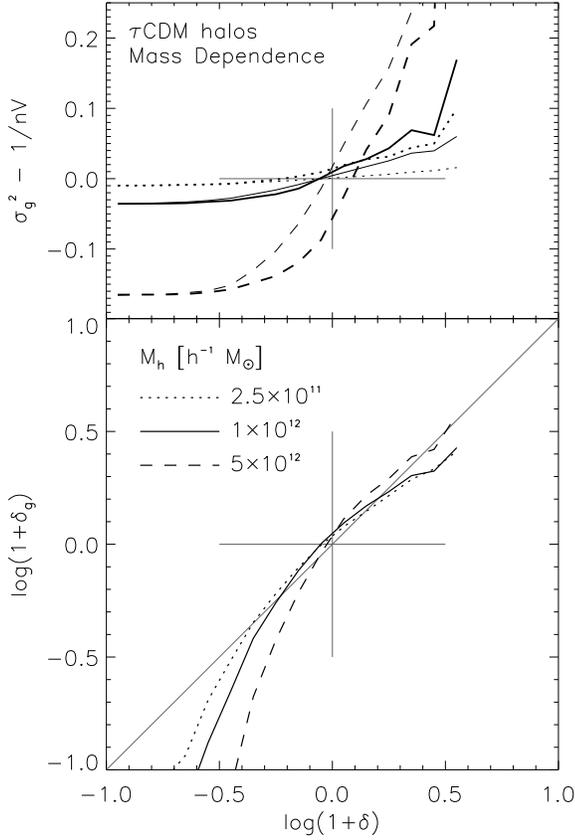,height=12truecm,width=8truecm}}
\caption{Mass dependence of the mean and scatter of the conditional
biasing relation for halos, for $\tau$CDM at $z=0$. The bottom panel
shows the mean conditional biasing relation for halos in our
simulations, selected with different mass thresholds. The top panel
shows the variance of the conditional biasing relation, minus the mean
expected variance due to shot noise. Bold lines indicate the
simulation results and light lines indicate the ``conditional
Poisson'' model discussed in the text. }
\label{fig:moments_mass}
\end{figure}

\begin{figure}
\centerline{
\psfig{file=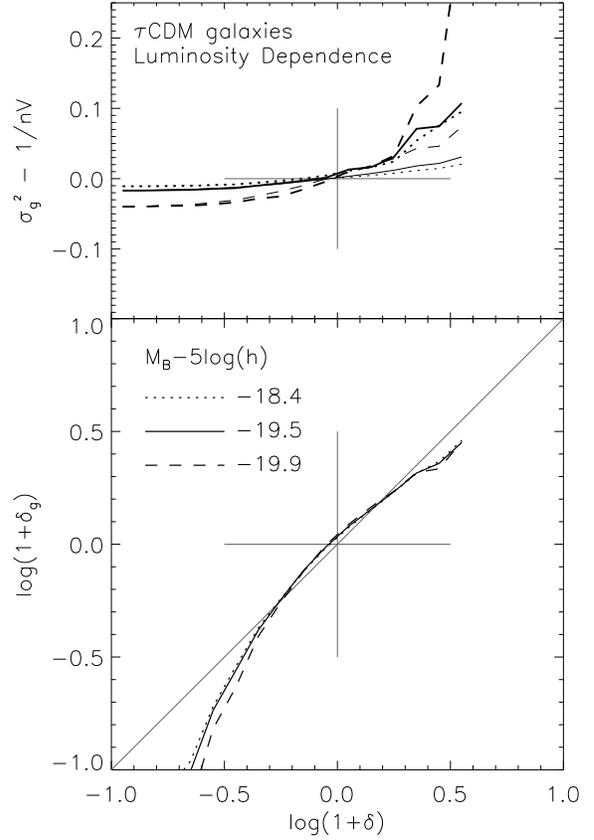,height=12truecm,width=8truecm}}
\caption{Same as figure~\protect\ref{fig:moments_mass}, for galaxies
selected with different B-band absolute magnitude thresholds. }
\label{fig:moments_lum}
\end{figure}

\begin{figure}
\centerline{
\psfig{file=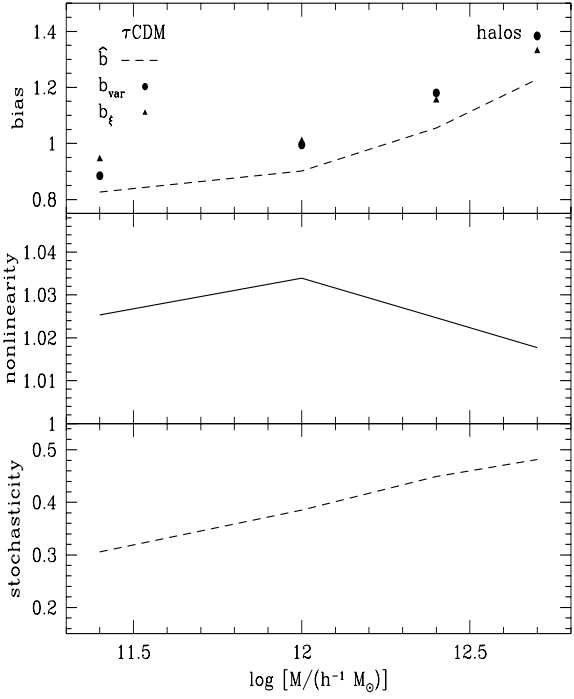,height=10truecm,width=8truecm}}
\centerline{\psfig{file=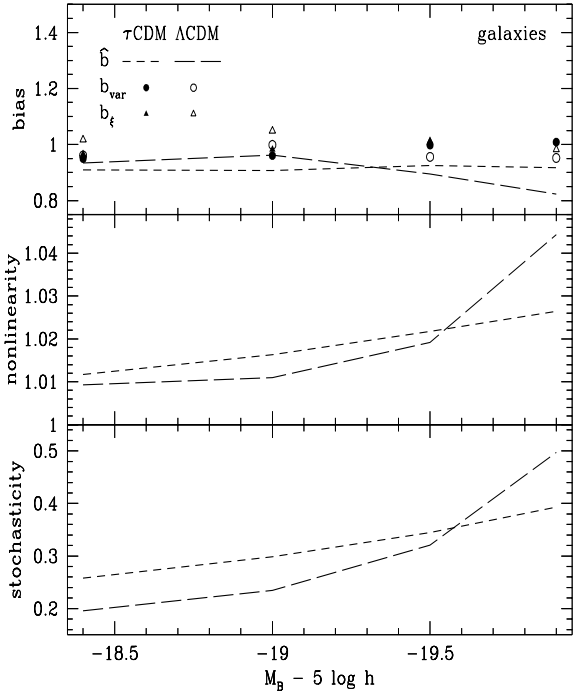,height=10truecm,width=8truecm}}
\caption{Biasing characteristics as a function of halo mass threshold and
galaxy magnitude limit.  The simulations are analyzed at $z=0$, with T8
smoothing. Top: mean biasing $\hat{b}$, $\bvar$, and $\bxi$.  Middle:
nonlinearity $\tilde{b}/ \hat{b}$.  Bottom: stochasticity $\sigma_{\rm b} /
\hat{b}$. }
\label{fig:bias_masslum}
\end{figure}

\begin{figure}
\centerline{\psfig{file=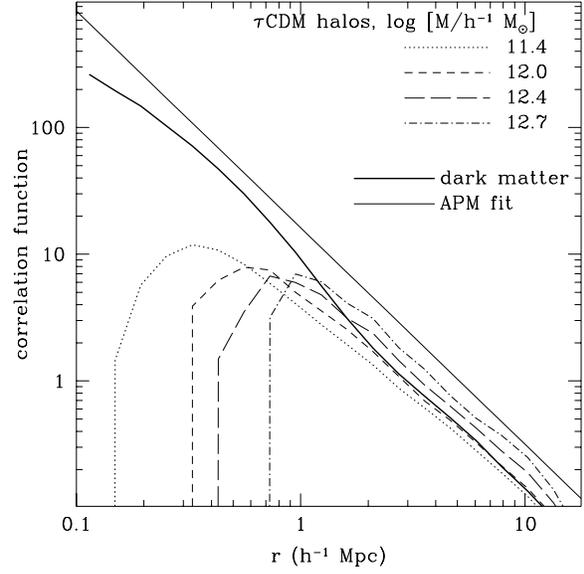,height=8truecm,width=8truecm}}
\caption{Biasing of halos as measured by the two-point auto-correlation
functions for different halo-mass thresholds at $z=0$ in the \taucdm\
simulations.  The bold solid line refers to the mass and the broken lines to
the halos.  The fit to the observed galaxy correlations in the APM survey is
shown for reference (light solid line). Results for the \lcdm\ simulations are
qualitatively similar. }
\label{fig:xi_mass}
\end{figure}

\begin{figure*}
\centerline{\psfig{file=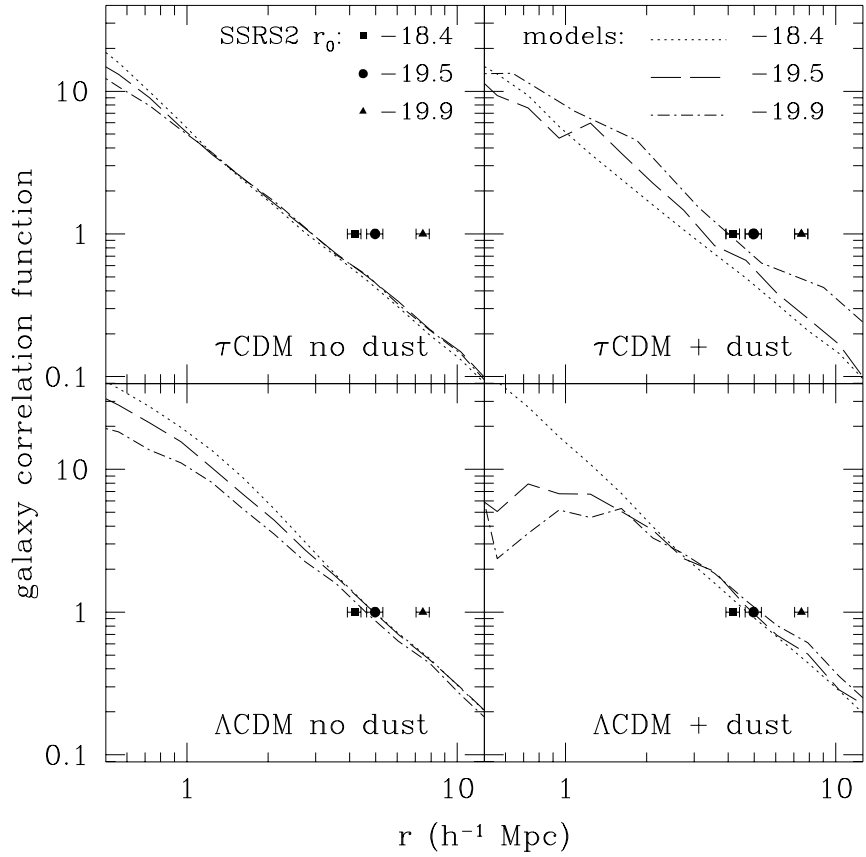,height=12truecm,width=12truecm}}
\caption{The auto-correlation function for galaxies (light broken lines) with
varying absolute magnitude limits at $z=0$ in the \taucdm\ simulation (top) and
\lcdm\ simulations (bottom). The right panels include a model for dust
extinction (see text). Symbols show the correlation length $r_0$ for galaxies
with varying absolute magnitude thresholds, obtained by
\protect\citeN{willmer:98} from the SSRS2 redshift survey. }
\label{fig:xig_lum}
\end{figure*}

\begin{figure*}
\centerline{\psfig{file=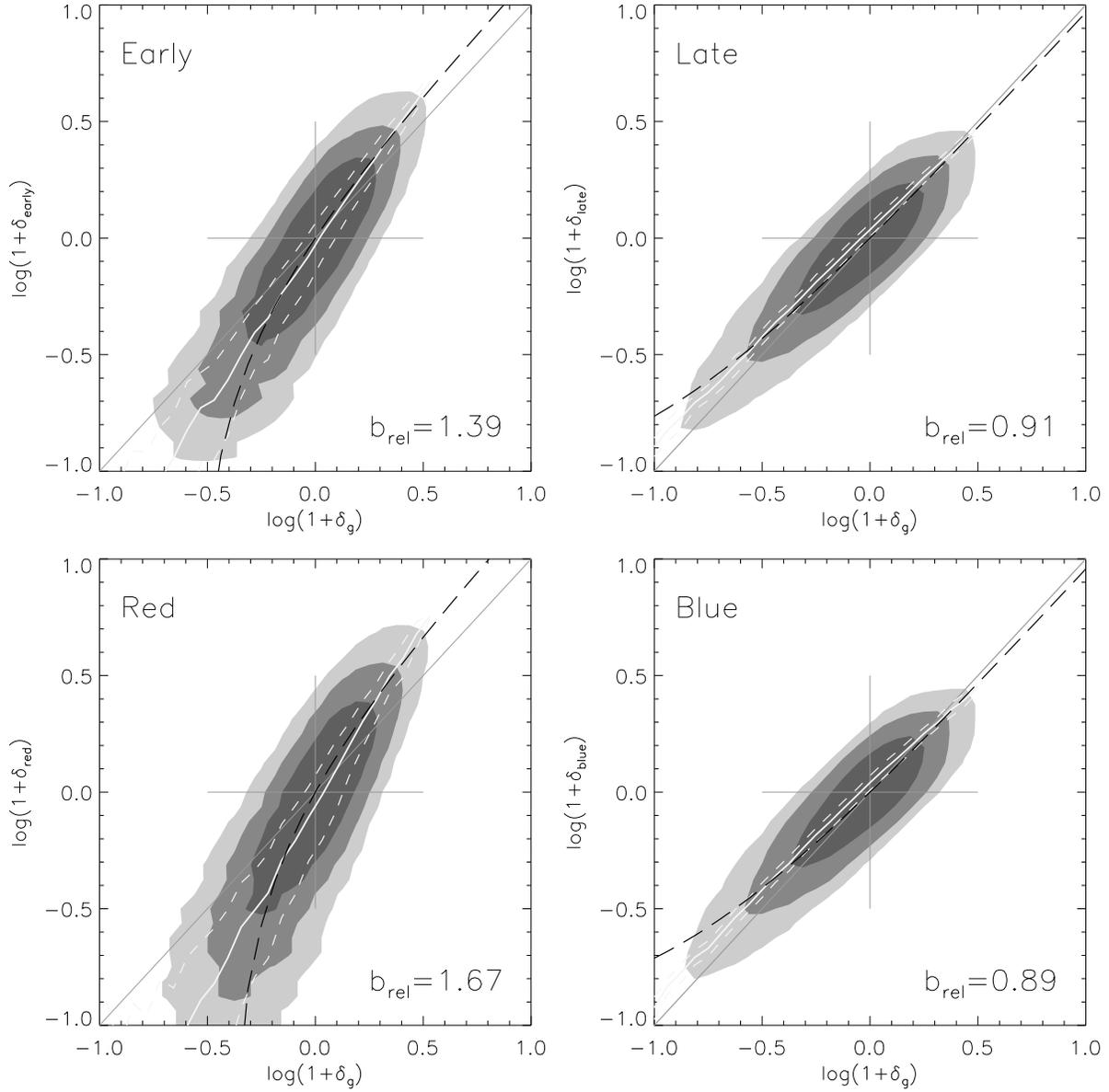,height=16truecm,width=16truecm}}
\caption{The joint distribution of the overdensity fields of type-selected
galaxies and \emph{all} galaxies, both with $M_B-5\log h \le -18.4$ and
smoothed with a top-hat window of radius $\Rs=8 \hmpc$, for the \taucdm\
simulations. The top panels represent galaxies selected by morphology
(``early'' corresponds roughly to E--S0, ``late'' to S--Irr Hubble types), and
the bottom panels represent galaxies selected according to B-V colour. The
contours represent approximately the 50 (dark grey), 80 (medium grey), and 98
(light grey) percentiles. The white lines show the mean conditional biasing
function $\av{\delta_{\rm early/late} | \delg}$ and the $1\sigma$ scatter about
it in equal bins of $\log \delg$. The black long-dashed line shows the linear
biasing approximation with $b=\hat{b}_{\rm rel}$.}
\label{fig:relbias}
\end{figure*}

\begin{figure*}
\centerline{\psfig{file=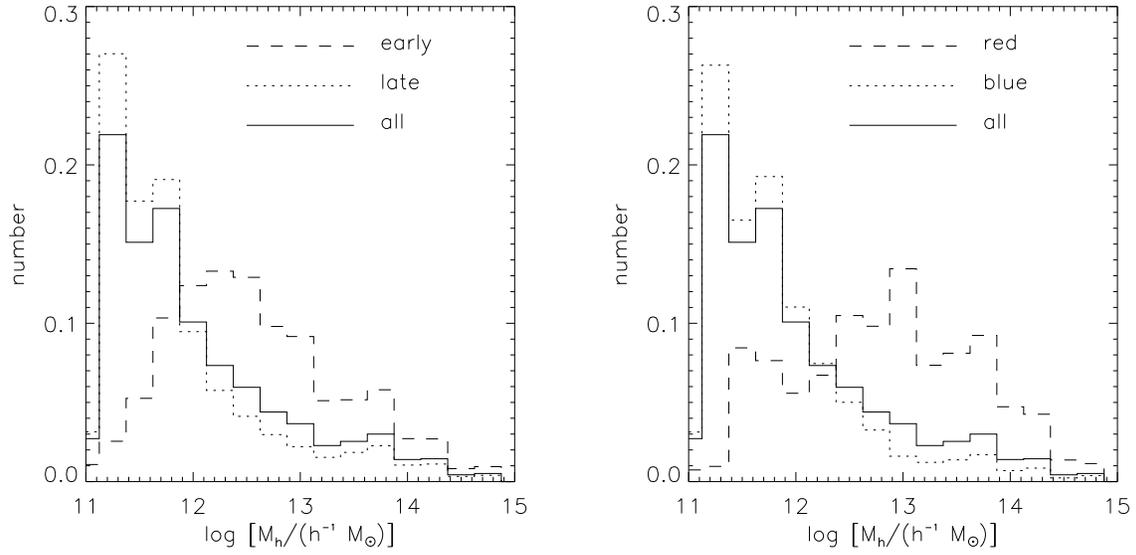,height=8truecm,width=16truecm}}
\caption{The distribution of host halo masses for galaxies selected according
to morphological type (left) and colour (right) and with $M_B - 5\log h \le
-18.4$. Solid lines show the distribution of host halo masses for all galaxies
above the magnitude limit, dotted lines show the distribution of masses for
late/blue galaxies, and dashed lines show the distribution for early/red
galaxies. Early-type and red galaxies dwell on average in more massive halos,
and hence are biased compared to the overall galaxy population. }
\label{fig:masshist}
\end{figure*}

\begin{figure}
\centerline{
\psfig{file=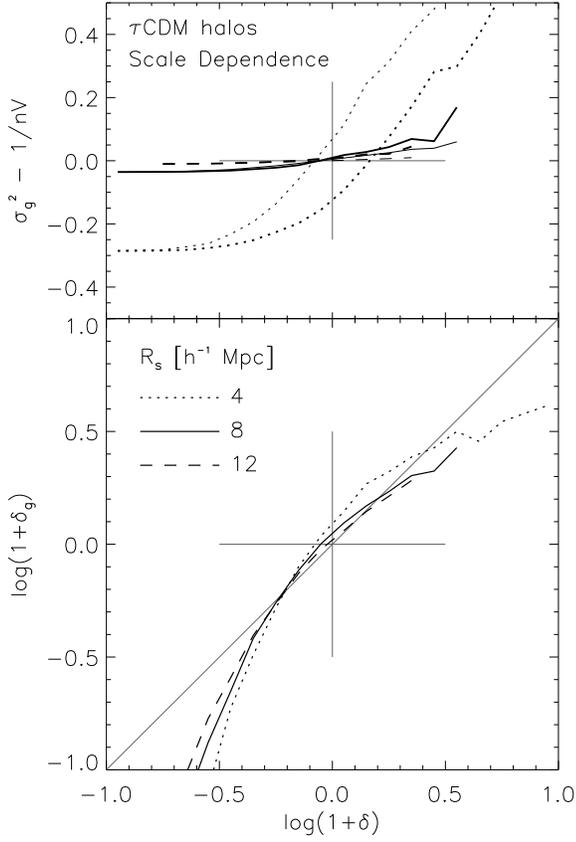,height=12truecm,width=8truecm}}
\caption{Scale dependence of the mean and scatter of the
conditional biasing relation for halos ($M \ge 10^{12} \hmsun$), for
$\tau$CDM at $z=0$. The quantities shown are as in
figure~\protect\ref{fig:moments_mass}. }
\label{fig:moments_scale_h}
\end{figure}

\begin{figure}
\centerline{
\psfig{file=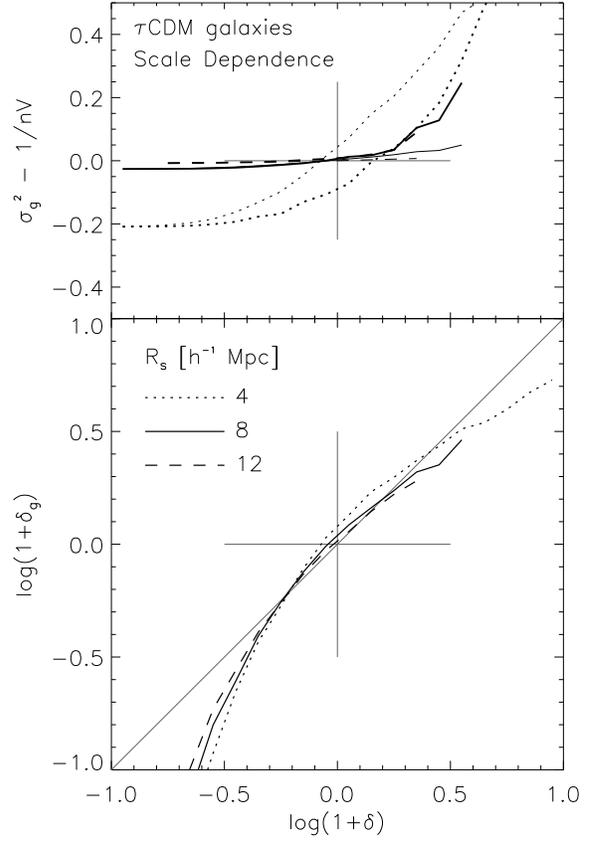,height=12truecm,width=8truecm}}
\caption{Same as figure~\protect\ref{fig:moments_scale_h}, for galaxies
with $M_B-5 \log h \le -19.5$. }
\label{fig:moments_scale_g}
\end{figure}

\begin{figure}
\centerline{
\psfig{file=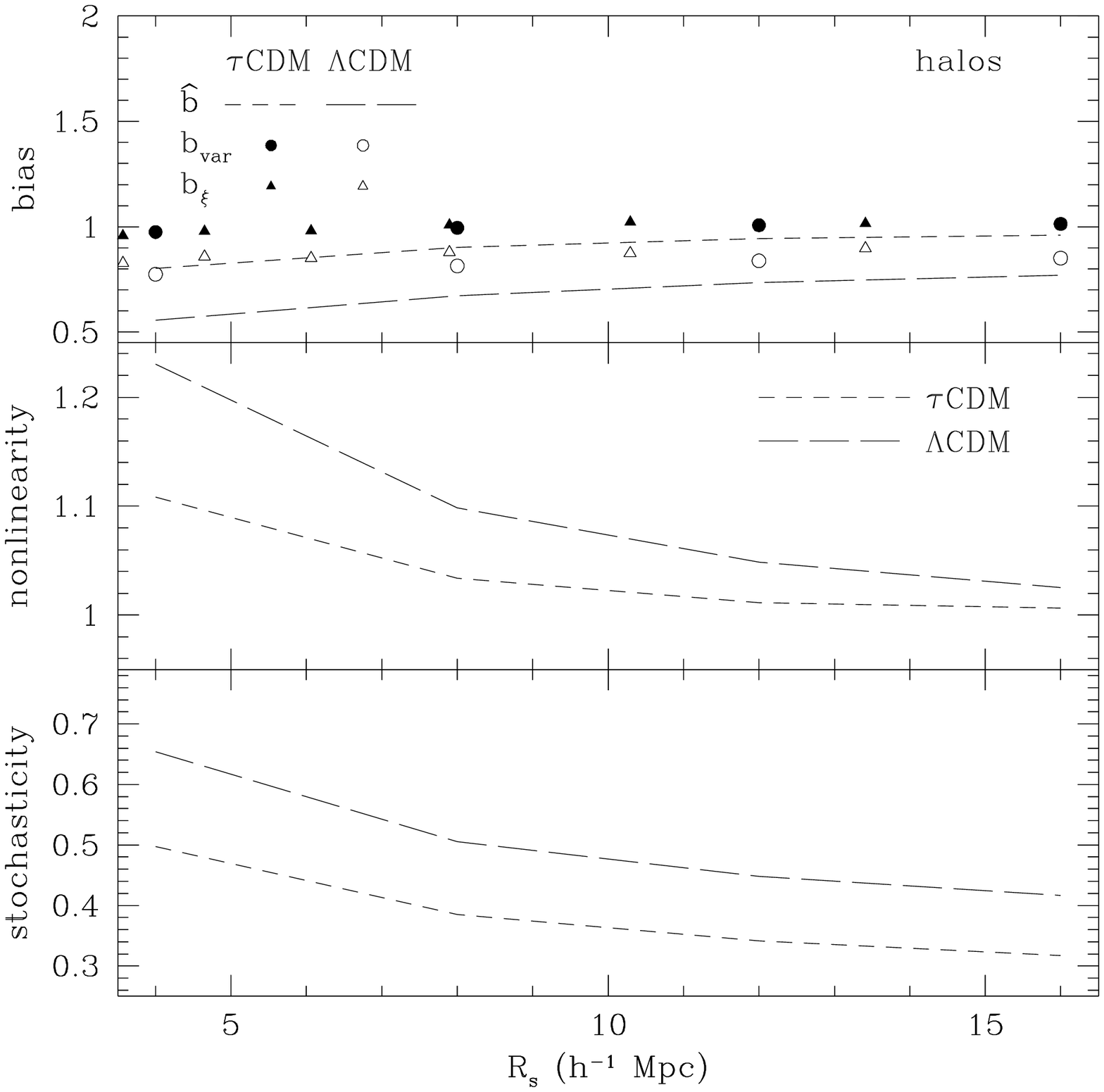,height=10truecm,width=8truecm}}
\centerline{
\psfig{file=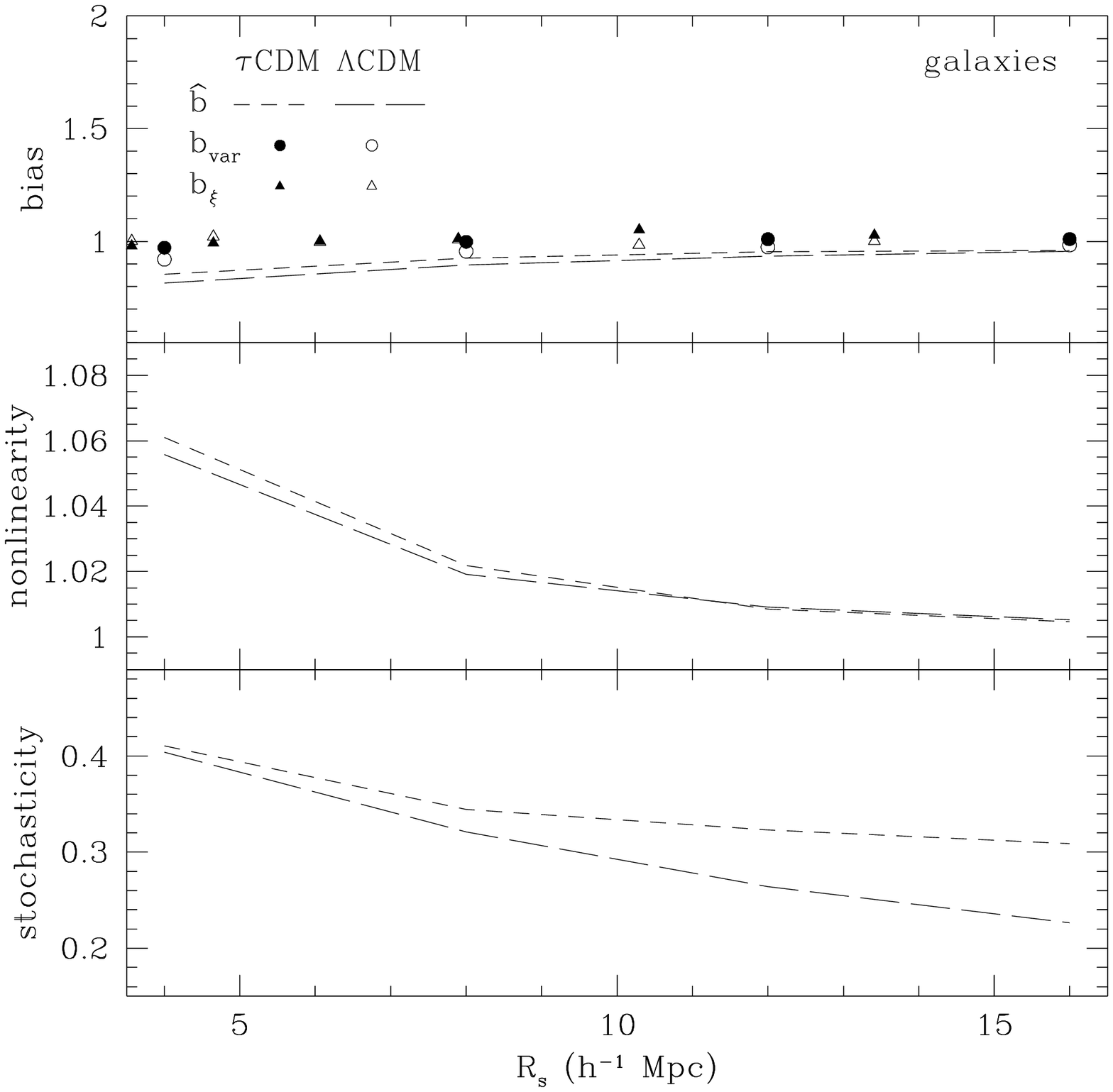,height=10truecm,width=8truecm}}
\caption{
Biasing characteristics as a function of smoothing scale for halos of
$M \ge 10^{12} \hmsun$ or galaxies with $M_B-5\log h \le -19.5$ at
$z=0$. Details are as in figure~\protect\ref{fig:bias_masslum}. }
\label{fig:bias_scale}
\end{figure}

We now return to the redshift evolution of biasing, which is the most
striking feature in figure~\ref{fig:dcontourh} and
\ref{fig:dcontourg}, and is of particular interest given the recent
observational developments at high redshift.  We use a fixed halo-mass
threshold ($M \ge 10^{12} \hmsun$) or luminosity threshold ($M_B-5\log
h \le -19.5$) and smoothing scale (T8). We already noted that the
shape of the biasing relation changes noticably, and in particular the
slope $\bhat$ increases dramatically with redshift. The rapid
evolution towards higher bias at high redshift has been noted in
several recent works
\cite{bagla:98,wechsler:98,colin:98,katz:98,blanton:99}. Here we
discuss several other interesting properties of the redshift evolution
of biasing of both halos and galaxies.

The \taucdm\ and \lcdm\ models, which showed very similar mass and
scale dependences at $z=0$, show quite different redshift
evolution. The evolution is more pronounced in the $\Omega_0=1$
\taucdm\ model than in the low-$\Omega_0$ \lcdm\ model, as
expected. This is because the mass and scale dependence at fixed
redshift are determined mainly by the shape of the power spectrum,
which is deliberately similar for these two models, while the redshift
evolution depends on the growth rate of clustering, which is quite
different in these two models.

In addition, we can see in figure~\ref{fig:dcontourh} and
\ref{fig:dcontourg} that the evolution of the biasing relation for
\emph{galaxies} is weaker than that for halos. This is because of the
redshift dependence of the efficiency of star formation and hence of
the relationship between halo mass and galaxy luminosity.  Halos are
typically denser at high redshift, and in these models we have
effectively assumed that the star formation rate is proportional to
the average halo density (see figure~8 and 9 of \citeN{kauffmann:98b}
and Eqn.~\ref{eqn:sf} of this paper). Therefore, when we select
galaxies with a fixed luminosity threshold, at high redshift we
include galaxies residing in smaller mass halos (and thus of lower
biasing). Note that this effect also reduces the difference in the
evolution of biasing within the two cosmological models, i.e., the
additional physics associated with star formation conspires to make
the redshift evolution of biasing \emph{less} discriminatory to
cosmology.

\begin{figure}
\centerline{
\psfig{file=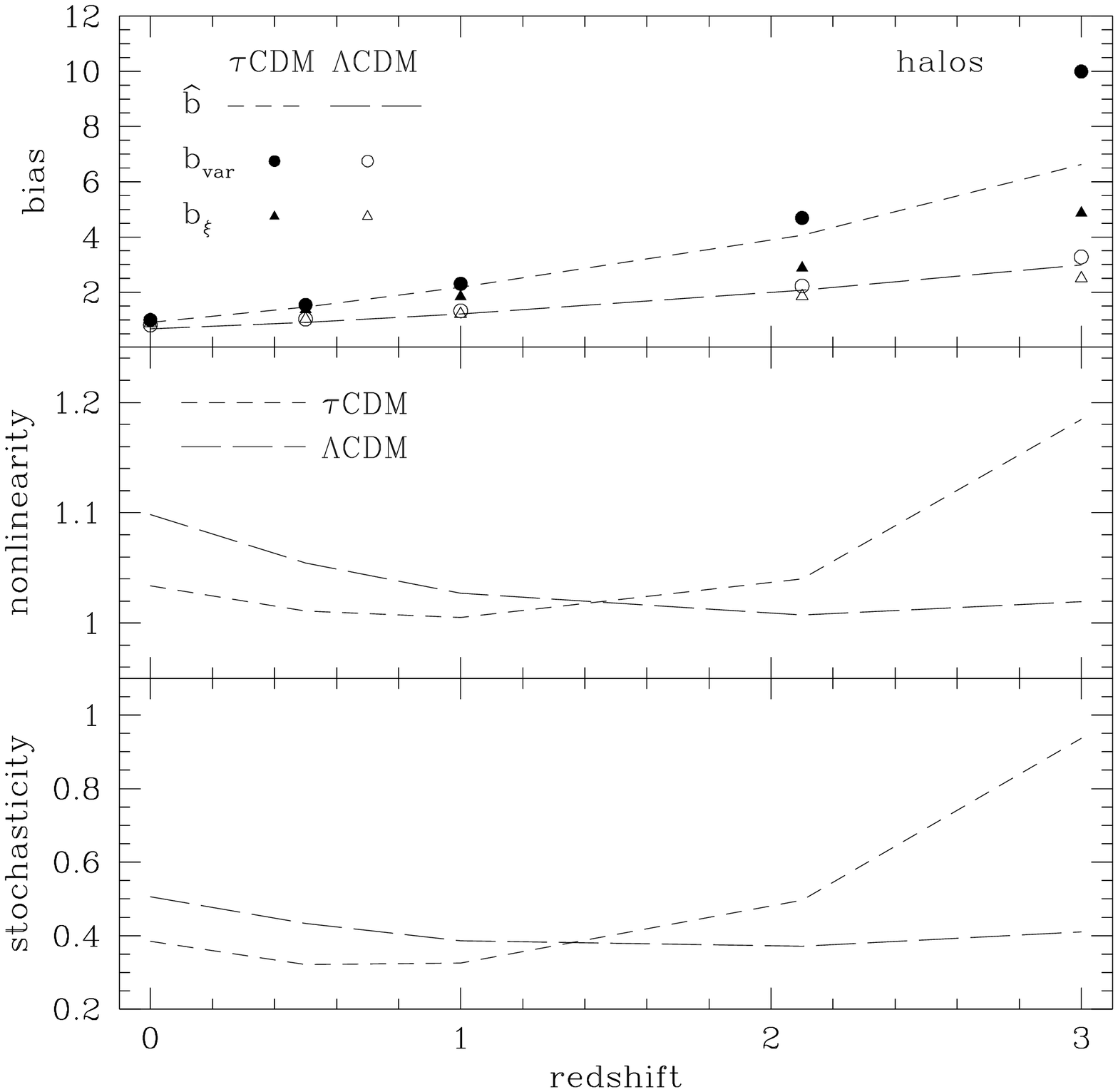,height=10truecm,width=8truecm}}
\centerline{\psfig{file=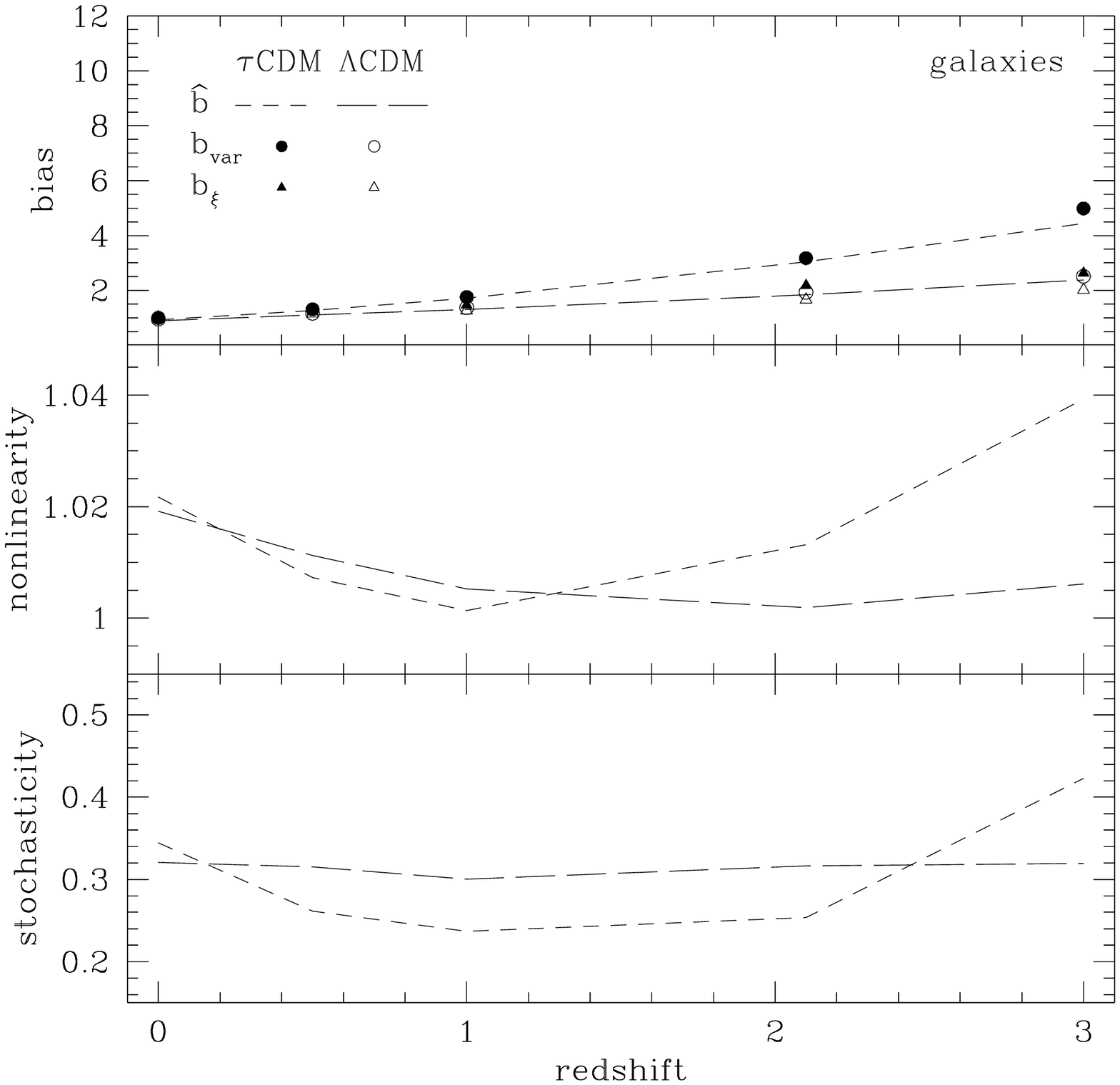,height=10truecm,width=8truecm}}
\caption{Biasing characteristics as a function of redshift.  The two
cosmological models are analyzed with T8 smoothing, for halos of $M
\ge 10^{12} \hmsun$ or galaxies with $M_B-5\log h \le -19.5$. Details
are as in figure~\protect\ref{fig:bias_masslum}. }
\label{fig:bias_z}
\end{figure}

Figure~\ref{fig:bias_z} summarizes the redshift evolution of the mean
bias, nonlinearity, and stochasticity parameters.  Note the increasing
difference between the mean biasing statistics, $\bhat$, $\bvar$ and
$\bxi$, with increasing redshift, for the \taucdm\ model; $\bhat$
differs from $\bxi$ by a factor of two at $z=3$. This difference
reflects the increase in stochasticity and non-linearity with redshift
in the \taucdm\ model, as seen in the lower panels of the figure.  The
increase in stochasticity in this model is dominated by the increase
in shot noise due to the decrease in number density of halos/galaxies
between $z=1$ and $3$.  The difference in $\bxi$ and $\bvar$ is due to
the stronger scale dependence at $z=3$ ($\bxi$ reflects the biasing at
a fixed scale, whereas $\bvar$ is an average over different
scales).  In the \lcdm\ simulations, the non-linearity and
stochasticity instead decrease with redshift, and therefore $\bhat$,
$\bvar$ and $\bxi$ agree within 30 percent at $z=3$.  Thus, the
evolution of the biasing parameters is characterized in the two models
by a minimum in the nonlinearity and stochasticity, which occurs at
$z\sim1$ in the \taucdm\ cosmology and $z\sim 3$ in the \lcdm\
cosmology.

\begin{figure*}
\centerline{\psfig{file=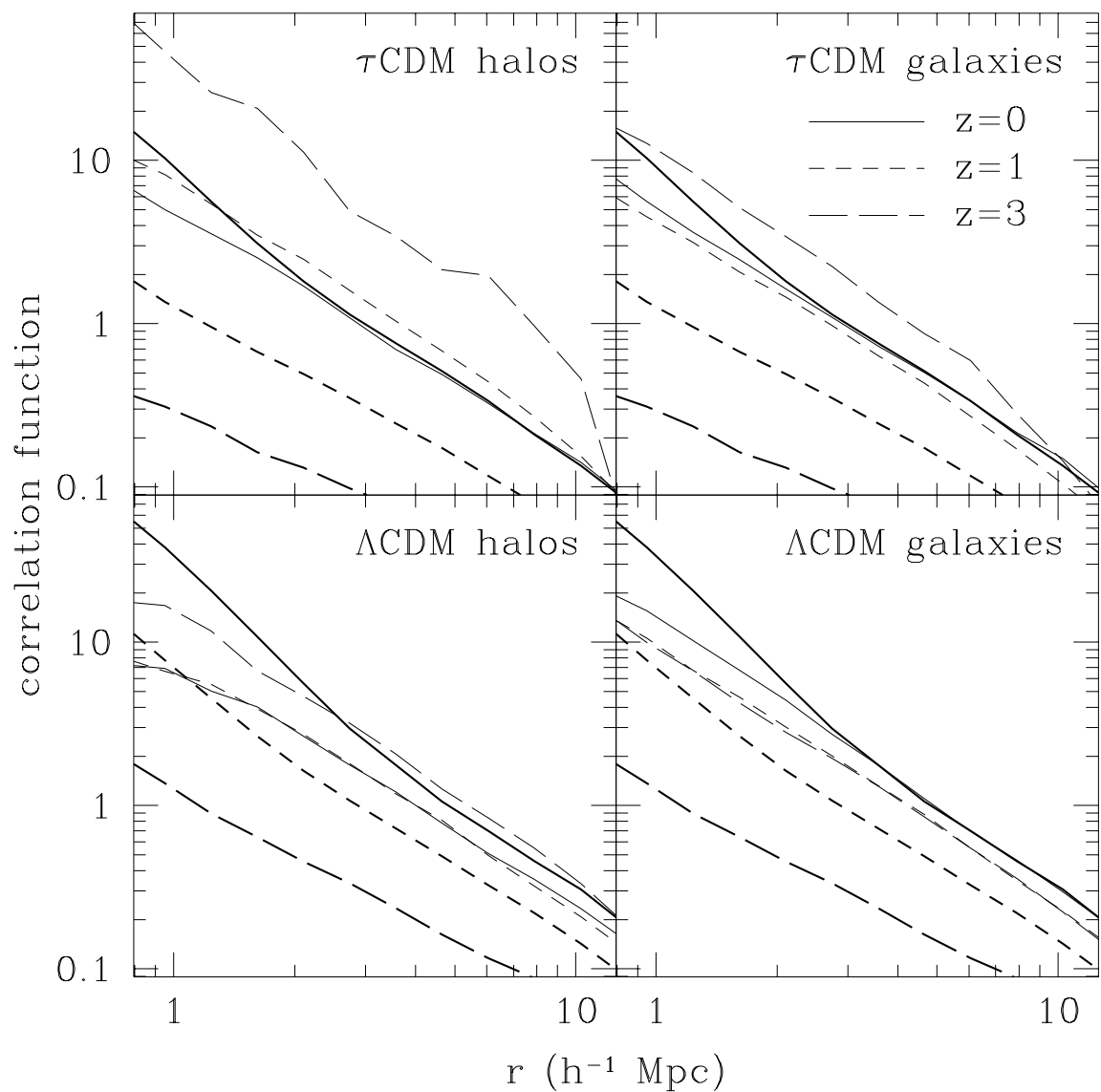,height=16truecm,width=16truecm}}
\caption{The auto-correlation function for halos ($M
\ge 10^{12} \hmsun$; left panels, light lines) 
or galaxies ($M_B-5\log h \le -19.5$; right panels, light lines) and
for the dark matter (bold lines) at different redshifts. }
\label{fig:xiz}
\end{figure*}

Some insight into the origin of this minimum may be gained by
examining the time evolution of the correlation functions of dark
matter, halos, and galaxies from $z=3$ to $z=0$, shown in
figure~\ref{fig:xiz}. The clustering amplitude of the dark matter
increases monotonically as time progresses. However, halos of a fixed
physical mass correspond to rarer peaks in the density field at higher
redshift, so the clustering amplitude of these objects
\emph{decreases} monotonically as time moves forward. The changing
mass scale corresponding to galaxies of a fixed B-band luminosity
discussed above leads to a rather different behaviour for the
magnitude-limited galaxies. In both models, the clustering amplitude
decreases as we move backwards in time from $z=0$ to $z=1$. It starts
increasing at a redshift of about $z=1$ for \taucdm\ and $z=3$ for
\lcdm. This effect is discussed in much more detail in
\citeN{kauffmann:98b}, who found that the magnitude of this ``dip'' in
the clustering amplitude or correlation length depends on the way in
which galaxies are selected as well as on the cosmology. In both
models investigated, the redshift at which the the correlation
amplitude begins to rise corresponds approximately with the minimum in
the non-linearity and stochasticity of the biasing relation noted
above. This correspondance is intriguing but its significance is not
immediately obvious.

\section{Comparison with Analytic Models of Biasing}
\label{sec:analytic}
Analytic models of biasing provide an important counterpart to
$N$-body simulations, which have limited resolution and volume. In
this section we evaluate several biasing models by comparing them with
the results of our simulations, focussing especially on the redshift
dependence. Numerous recent papers
\cite{matarrese:97,moscardini:98,adelberger:98,coles:98,mmw:99,magliocchetti:99}
have made use of versions of these analytic models to make predictions
about galaxy clustering at high redshift and to interpret
high-redshift observations.

We first investigate the Mo \& White model, summarized in
Section~\ref{sec:formalism:models:mw}. In their original paper,
\citeN{mw} showed that their model agreed well with the results of
simulations with scale-free power spectra.  It is useful to
investigate their model in the context of a more realistic power
spectrum, and to show results in terms of the physical mass scales and
redshifts that we expect to correspond to observable galaxies. For
technical convenience, we display the biasing relation as a function
of mass \emph{threshold}, while MW originally provided predictions for
halos of a specific mass.  To compute the MW prediction for all masses
above a threshold, we simply perform a weighted average over the
original expression (MW eqn. 19), using the Press-Schechter formula
for the number density of halos of a given mass \cite{ps:74}.

\begin{figure}
\centerline{
\psfig{file=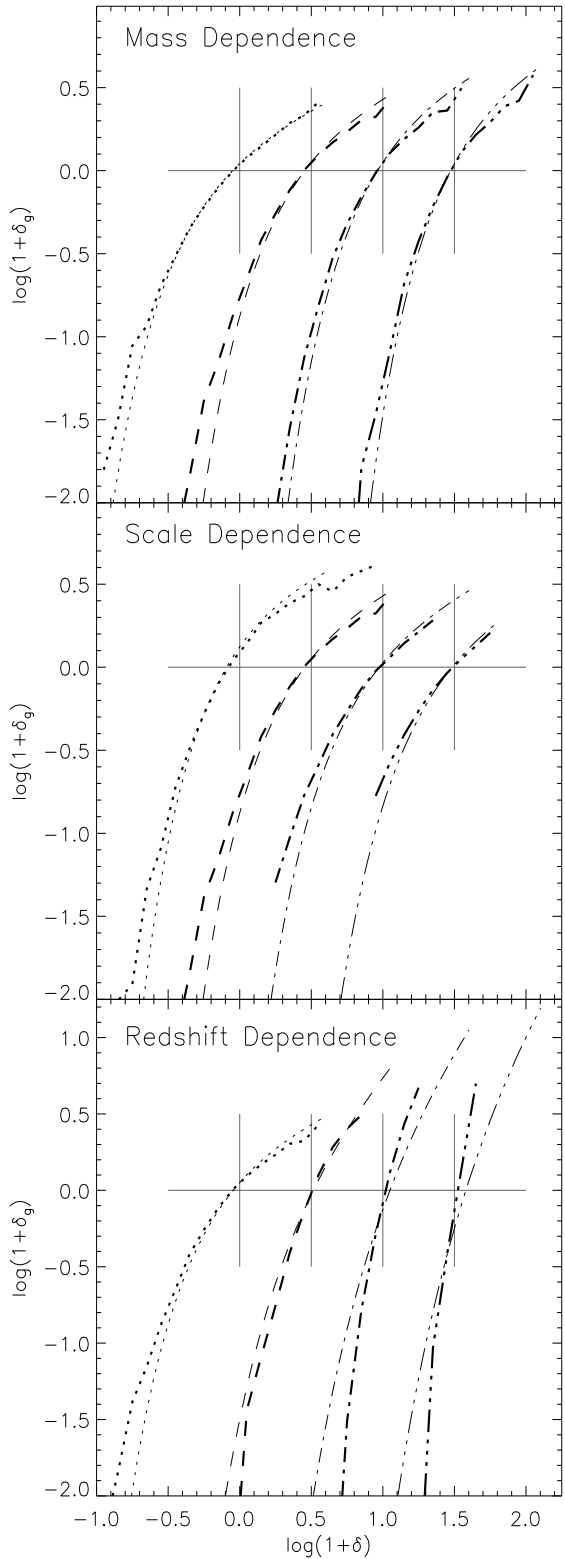,height=22truecm,width=8truecm}}
\caption{Comparison of the predictions of the MW model (light lines) with the 
simulation results (bold lines) for the mean biasing relation. The
curves in each panel have been offset by 0.5 dex for clarity. The top
panel shows the results for different mass thresholds ($\log
[M/(\hmsun)] = 11.4, 12.0, 12.4, 12.7$ from left to right), the middle
panel shows different smoothing scales ($R_s = 4, 8, 12, 16 \hmpc$
from left to right) and the bottom panel shows different redshifts
($z=0$, 1, 2, 3 from left to right).}
\label{fig:mwcomp}
\end{figure}

Figure~\ref{fig:mwcomp} shows the comparison of the MW predictions for
the mean biasing relation with the simulation results (\taucdm) for
halos at different mass thresholds, smoothing scales, and
redshifts. Apparently the MW model does well at reproducing the
changing shape of the biasing relation with mass, scale, and to a
lesser extent, redshift.  At $z=0$, the MW model tends to slightly
underpredict the halo overdensity in regions of very low matter
overdensity. At larger redshifts, the MW model fails to follow the
rapid growth of the slope $\bhat$ with redshift.  The progressively
larger deviation of the MW predictions from the simulation results
with increasing redshift is probably the result of inaccuracies in the
underlying Press-Schechter and extended Press-Schechter
approximations, which are known to have a similar redshift dependence
\cite{slkd}. However, the MW model does correctly predict the
qualitative features of the evolution, including both the change in
shape and the progression towards higher biasing.

\begin{figure}
\centerline{
\psfig{file=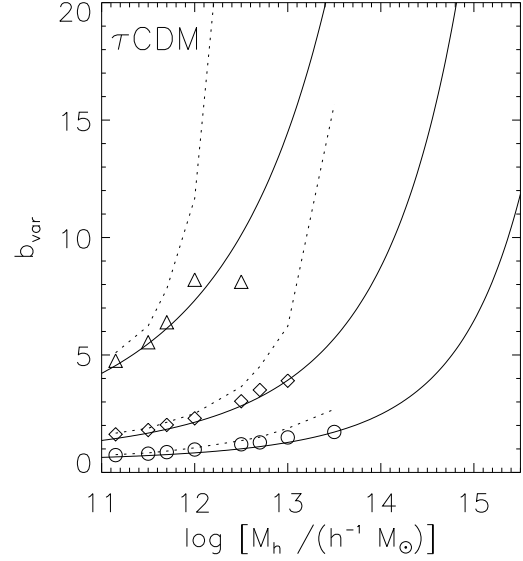,height=8truecm,width=8truecm}}
\caption{Mass dependence of halo biasing ($\bvar$) at $z=0$, $z=1$,
and $z=3$. Solid lines are the predictions of the MW model. Dotted
lines are results from the \taucdm\ simulations without correction for
shot-noise. Symbols are results from the simulations with a standard
Poisson correction for shot-noise. Circles and the lowest set of lines
are for $z=0$, diamonds and the middle set of lines for $z=1$, and
triangles and the highest lines are for $z=3$. }
\label{fig:bias_mass_mw}
\end{figure}

As we mentioned in Section~\ref{sec:formalism:models:mw}, when the
overdensity within the smoothing window is low compared to the
critical overdensity for collapse at a particular epoch, the MW model
reduces to a very simple expression for the linear bias factor b(M, z)
as a function of halo mass and redshift (MW eqn.~20, our
eqn.~\ref{eqn:mw}). Figure~\ref{fig:bias_mass_mw} shows the mass
dependence of the halo biasing parameter $\bvar$ at several redshifts
for the simplified MW model and the \taucdm\ simulations. Even this
simple expression agrees remarkably well with the mass and redshift
dependence of biasing of halos in our simulations. The mass range of
halos that can be usefully studied in our simulations is limited on
the low mass end by our resolution ($M_{\rm res} = 2.0 \times 10^{11}
M_{\odot}$) and on the high mass end by shot noise. In the figure, we
show the simulation results both with and without a standard Poisson
correction for shot noise. Particularly for high mass halos, the
sampling is far from Poisson and the shot noise correction is likely
to be inaccurate. Another interesting thing to note is that the mass
dependence of biasing over scales comparable to the size of galactic
halos ($\sim 10^{11}-10^{12} \hmsun$) is much stronger at high
redshift than at $z=0$. This is because of the dramatic decrease of
the non-linear clustering mass $M_{*}$ with redshift (see
figure~\ref{fig:mstar}).

\begin{figure}
\centerline{
\psfig{file=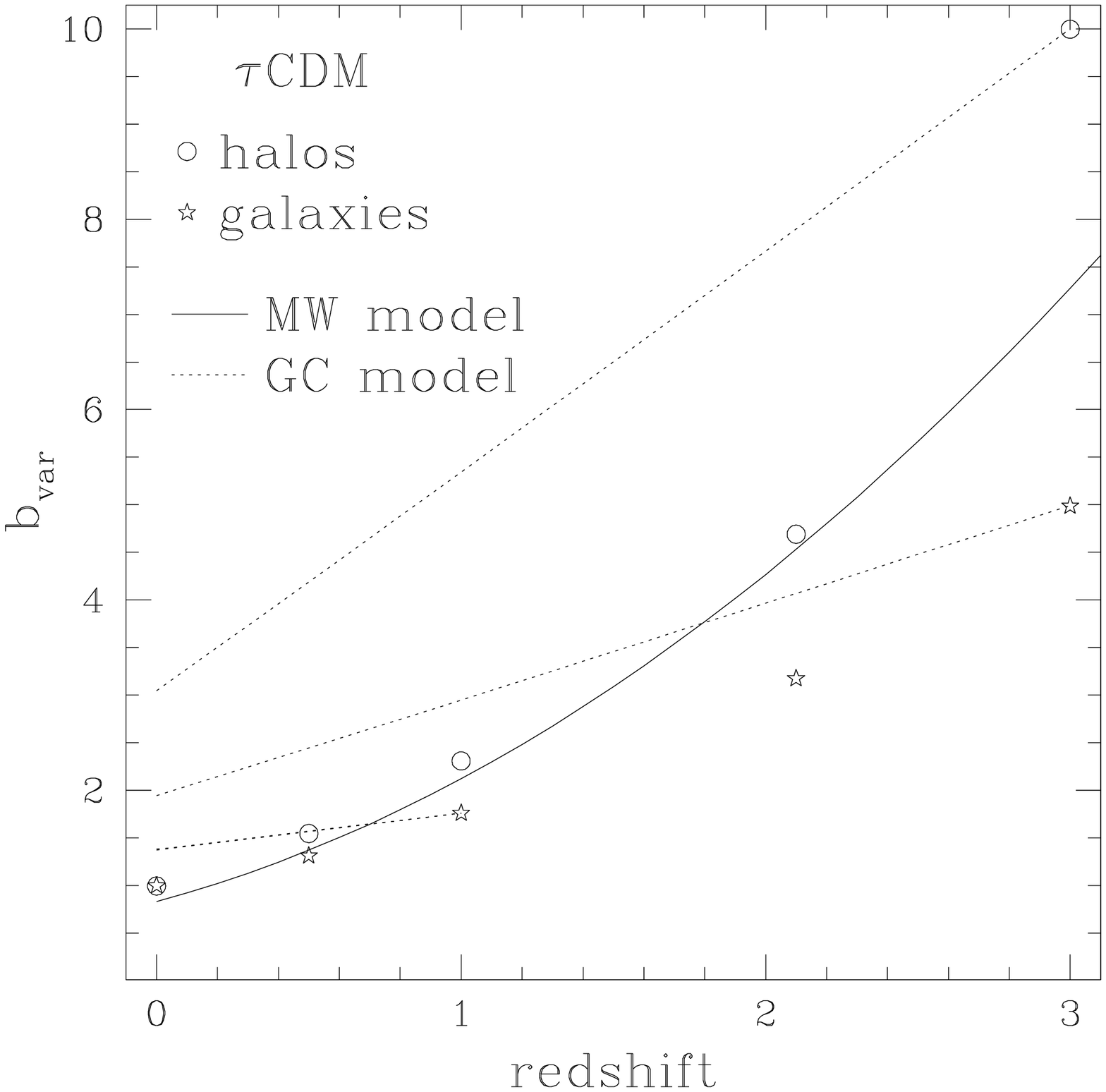,height=8truecm,width=8truecm}}
\centerline{
\psfig{file=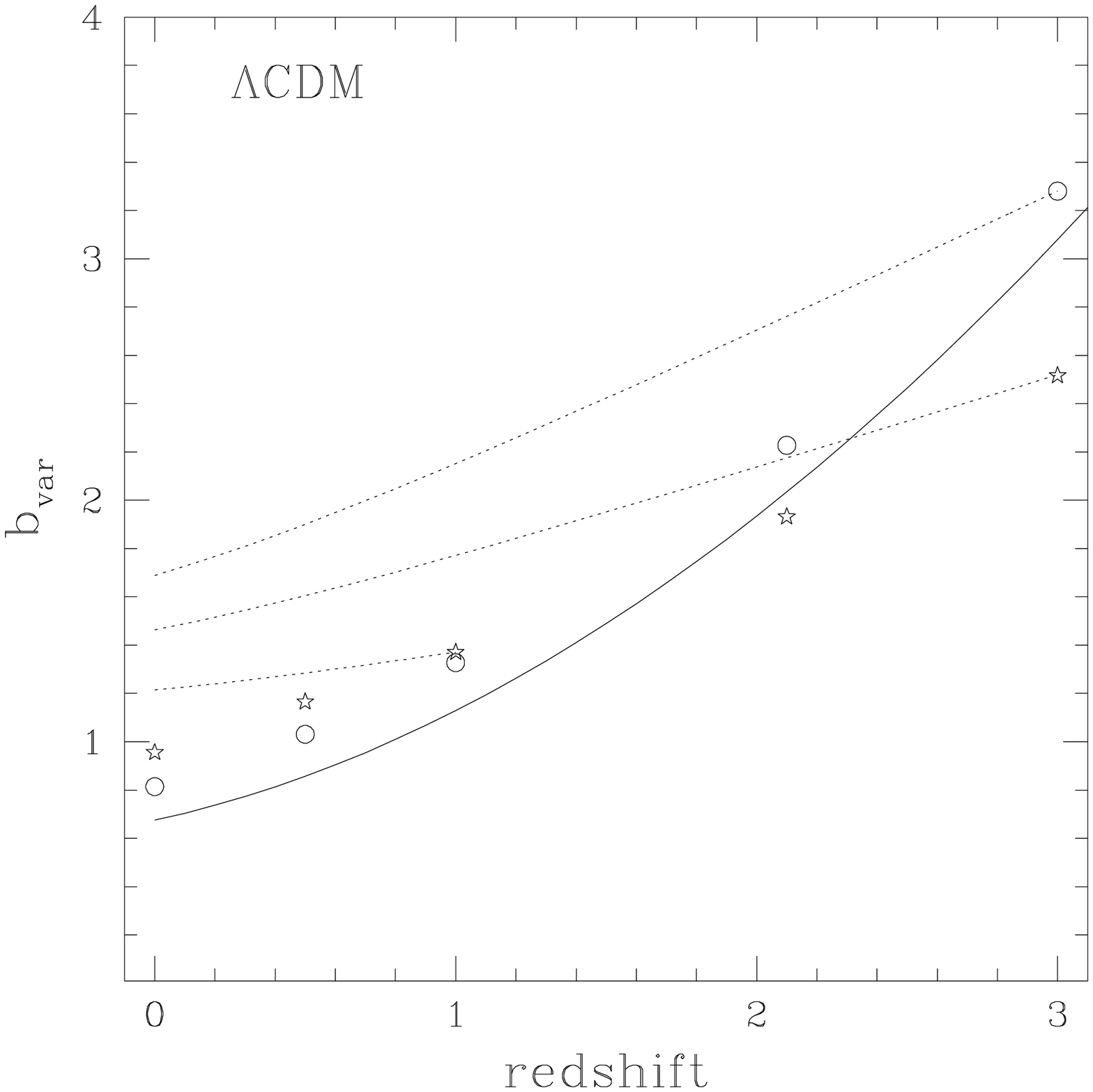,height=8truecm,width=8truecm}}
\caption{Redshift evolution of biasing ($\bvar$) of galaxies
($M_B-5\log h \le -19.5$; stars) and halos ($M \ge 10^{12} \hmsun$;
dots) in the simulations, compared with analytic models. The light
solid lines show the predictions of the MW model, with the same mass
threshold as the simulation halos. The dotted lines show the
predictions of the galaxy-conserving model, for initial bias values
typical of halos or galaxies at $z=3$ or galaxies at $z=1$ in our
simulations (see text for details). }
\label{fig:analytic_bias_z}
\end{figure}

The time evolution of $\bvar$ for halos ($M > 10^{12} \hmsun$) and
galaxies ($M_B-5\log h \le -19.5$) is compared with the predictions of
the models in figure~\ref{fig:analytic_bias_z}. One can clearly see
the slower evolution of galaxies versus halos in the simulations,
discussed in Section~\ref{sec:results:redshift}. We show the MW model
(eqn.~\ref{eqn:mw}) for the corresponding mass threshold. The MW model
gives a good qualitative description of the evolution of {\it halo}
biasing, as noted before.

There is \emph{no} mass threshold that gives a good description of the
evolution of the {\it galaxy} biasing, because of the differing merger
timescales of galaxies and halos, and because of the effects mentioned
earlier concerning the changing mass-to-light connection as a function
of redshift.

We also show the galaxy-conserving (GC) model
\cite{fry:96,tegmark-peebles:98} described in
Section~\ref{sec:formalism:models:fry}. We start at high redshift
($z=3$ or $z=1$) with the values of $\bvar$ and the linear correlation
coefficient $r$ as measured from the simulation galaxies or halos, and
propagate the bias to $z=0$ using eqn.~\ref{eqn:tp}. The evolution
predicted by the GC models is much weaker than that predicted by the
MW models and observed in the simulations. This clearly suggests that
merging, which is not included in the GC model, is an important
process in the evolution of clustering over this redshift range, even
in a low-$\Omega$ Universe. We might expect that merging would be less
important for galaxies than halos, and indeed the weak evolution
according to the GC model is slightly closer to the biasing evolution
for the galaxy population, but it still significantly underpredicts
the rate of evolution.

\section{Summary and Conclusions}
\label{sec:summary}
We have performed a study of biasing of dark-matter halos and galaxies
in cosmological simulations of \lcdm\ ($\Omega=0.3$) and \taucdm\
($\Omega=1$), explicitly treating and quantifying the mean biasing as
well as non-linearity and stochasticity in the biasing relation. We
conclude that the CDM-based hierarchical structure formation scenario
predicts that biasing has a moderate degree of nonlinearity and
stochasticity, and it depends on mass, type, scale and especially
redshift.

The mean biasing function is always of the following characteristic
shape. In the underdense regime ($-1<\delta<0$), $b(\delta)$ vanishes
near $\delta\gsim -1$, then rises sharply towards $\delg\simeq
\delta=0$ with a slope steeper than unity.  In the overdense regime
($\delta>0$), the behavior is fairly linear, and the effective slope
of $b(\delta)$ is either larger or smaller than unity, representing
biasing or anti-biasing.  The nonlinearity at $8\hmpc$ is at the level
of a few to $\sim$10 percent. The stochasticity is typically at the
level of a few tens of percent.

For halos, we find that the mean biasing increases gradually with
\emph{mass}, as does the stochasticity (to a large extent due to an
increasing shot noise contribution), while the non-linearity is nearly
the same for all masses.  At $z=0$, the biasing relation for galaxies
in our simulations is similar to that for $\ga 10^{12} \hmsun$ halos.
This reflects the fact that our halos of $10^{12} \hmsun$ contain
\emph{on average} one bright galaxy per halo, but is still somewhat
surprising given that the halos do contain different numbers of
galaxies with various luminosities.  We find that the biasing is
nearly independent of the luminosity of the galaxies selected, in
contradiction with some observational results, but in keeping with our
finding of weak mass dependence of halo biasing over the mass scales
typical of galaxy-sized halos at $z=0$.

This apparent contradiction may be resolved by considering dust
extinction.  We show that including the effects of dust effectively
increases the mass-to-light ratio, so that $L>L_{*}$ galaxies are
found in larger mass halos where the biasing dependence on mass is
more pronounced.  This leads to a luminosity dependent biasing that is
qualitatively similar to that observed, although neither of the
cosmological models considered here fully succeed in simultaneously
reproducing the amplitude of the observed correlation function and the
detailed luminosity dependence of galaxy clustering. We suspect that a
cosmology with an intermediate value of $\Omega_0\sim 0.5$ might give
better results. However, the observational results are still somewhat
controversial, and the modelling of dust extinction is highly
uncertain. Since this appears to have the potential to be a rather
powerful constraint, the relative biasing of galaxies of different
luminosities should be measured more accurately using larger
samples. Ideally, this should be investigated using a K-band selected
sample which will decrease the uncertainties connected with dust
extinction.

We find that galaxy biasing depends fairly strongly on morphological
\emph{type} and \emph{color}, with early type and red galaxies being
respectively about 1.3 to 1.8 times more biased than late type or blue
galaxies. These results are in good agreement with current
observational estimates. More detailed studies of \emph{relative}
biasing will be an important application of forthcoming large redshift
surveys such as 2dF, SDSS, and 2MASS.

Over the range of smoothing \emph{scales} from $4$ to $16 \hmpc$, the
mean biasing increases weakly with smoothing scale, similarly for
halos and galaxies.

The non-linearity and stochasticity decrease with increasing smoothing
scale, so that the linear deterministic biasing approximation is
approached for very large smoothing scales, as expected. On scales
larger than a few Mpc, the clustering of bright galaxies in our
simulations is fairly robust to differing assumptions about the
details of galaxy formation. On smaller scales, halos become strongly
anti-correlated because of exclusion effects. This implies that in
order to reproduce the observed unbroken powerlaw correlation function
of galaxies on these scales, there must be a significant number of
galaxy pairs that cohabit the same halo. The biasing of our simulated
galaxies on these scales is highly sensitive to the details of the
astrophysical recipes, e.g., supernovae feedback, dust extinction,
etc.

The mean biasing of halos of a fixed mass and smoothing scale
increases dramatically with \emph{redshift}, by a factor of 5-10 for
\taucdm\ and a factor of $\sim 3$ for \lcdm\ from $z=0$ to $z=3$. In
both models, the nonlinearity of the biasing relation evolves with
redshift, decreasing to a minimum value very close to 1.0 (pure linear
bias) at $z\sim1$ for \taucdm\ and $z\sim3$ for \lcdm, then increasing
again at higher redshift.  The redshift dependence of biasing of
galaxies with a fixed luminosity threshold differs significantly from
that of halos with a fixed mass threshold. This is because, due to our
standard star-formation recipe, galaxies of a given luminosity are
hosted at higher redshift by halos of smaller masses and thus of
weaker biasing.

We have compared the results of the numerical simulations with
analytic models for biasing. The MW model provides a good description
of the changing shape of the biasing relation for halos as a function
of mass threshold, smoothing scale, and more qualitatively,
redshift. Even the simplified linear version of the MW model
(Eqn.~\ref{eqn:mw}) provides a fairly accurate description of the mass
dependence of mean halo biasing and the redshift evolution of halos of
a fixed mass. The evolution of \emph{galaxy} biasing may differ
substantially depending on the efficiency of star formation and how it
varies with redshift. Galaxy-conserving models (e.g. \citeNP{fry:96};
\citeN{tegmark-peebles:98}) substantially underpredict the evolution
rate of halo \emph{and} galaxy biasing at all epochs, strongly
suggesting that \emph{merging} is a crucial ingredient in the
evolution of biasing over this redshift range.

Our results are relevant to the attempts to measure the cosmological
density parameter using galaxy densities, e.g., via redshift
distortions or via a comparison to streaming velocities. Under the
simplified assumption of linear and deterministic biasing, these
attempts lead to a measure of the quantity $\beta \equiv
\Omega^{0.6}/b$.  Dekel \& Lahav (1998) have proposed that deviations
from the linear deterministic biasing ansatz may partially reconcile
the differing values of $\beta$ obtained by the different methods. Our
results suggest that the expected levels of nonlinearity and
stochasticity on the relevant scales, as predicted by simulations with
realistic models of galaxy formation, would lead to modest differences
in the various measures of $\beta$, on the order of 20--30 percent.
We suggest that the strong type dependence of biasing that is present
both in our simulations and in observations may also contribute to
some of the discrepancies in the results from different surveys, which
include different mixtures of galaxy types.  This should be
investigated further by applying these methods to detailed mock
catalogs extracted from simulations similar to those used here.

Our results also suggest several cautions that should be applied to
the numerous recent attempts to draw conclusions from the redshift
evolution of galaxy clustering. This may differ significantly from the
redshift evolution of halo clustering because of differences in
merging rates and a time dependence of the efficiency of star
formation. Moreover, if different types of galaxy are selected at
different redshifts, their biases may differ, giving a warped view of
the actual redshift evolution. Finally, we find that different
statistics for measuring the mean biasing may differ by as much as a
factor of 2, and the disagreement of various statistics changes with
epoch in accord with the changing importance of non-linearity,
stochasticity, and scale dependence. These statistics are often
treated as equivalent in the literature, an unfortunate outgrowth of
the linear deterministic biasing ansatz.

Our results are in general agreement both with other recent work using
different methods and with what we know about the scale, type and
redshift dependence of galaxy biasing from observations. Further
studies using these techniques, along with new data from forthcoming
large redshift surveys will no doubt lead to improved measurements of
the cosmological parameters using galaxy-based methods, and in
addition to a better understanding of the process of galaxy formation
and evolution.

\section*{Acknowledgements} 
We thank Ofer Lahav, Michael Strauss, Idit Zehavi, and Adi Nusser for
useful discussions. This research was supported by the US-Israel
Binational Science Foundation grants 95-00330 and 98-00217, and by the
Israel Science Foundation grants 950/95 and 546/98. The GIF
simulations were carried out at the Computer Center of the Max Planck
Society in Garching, Germany and at the Edinburgh Parallel Computing
Center, Scotland using codes kindly made available by the Virgo
Supercomputing Consortium. We thank J\"{o}rg Colberg and Adrian
Jenkins for carrying out the simulations.

\bibliographystyle{mnras} 
\bibliography{mnrasmnemonic,bgif}
\clearpage
\clearpage
\clearpage
\clearpage
\clearpage
\clearpage
\clearpage
\clearpage
\clearpage
\clearpage
\clearpage
\clearpage
\clearpage
\clearpage
\end{document}